\documentclass[pdftex, a4paper, 12pt]{article}%
\usepackage[square, numbers, comma, sort&compress]{natbib}
\usepackage[english]{babel}
\usepackage{booktabs}
\usepackage{subfigure} 
\usepackage{graphicx}
\usepackage{natbib} 
\usepackage{apalike} 
\usepackage{authblk}
\usepackage[top=1.2in, left=1.2in, right=1.2in]{geometry}
\usepackage{amsmath,amsfonts,amssymb}
\usepackage[framemethod=tikz]{mdframed}
\usepackage{hyperref}
\usepackage{rotating}
\usepackage{float}

\providecommand{\U}[1]{\protect \rule{.1in}{.1in}}

\hypersetup{colorlinks = true, urlcolor=black, citecolor = black, linkcolor= black}
\definecolor{MidnightBlue}{rgb}{0, 0, 0.551}
\definecolor{Gray}{gray}{0.85}
\setcitestyle{authoryear,open={(},close={)}}

\newcommand*\samethanks[1][\value{footnote}]{\footnotemark[#1]}

\begin{document}

\title{{\Large A Wavelet Method for Panel Models with Jump Discontinuities in the Parameters}}

\author{
Bada O.\thanks{Deutsche Telekom AG, 53227 Bonn, Germany. The opinions expressed in this article are the author's own and do not necessarily represent the views of Deutsche Telekom AG.}
\thanks{Corresponding authors:\\Oualid Bada (oualid.bada@telecom.de) and
  Dominik Liebl (dliebl@uni-bonn.de)},
Kneip A.\thanks{Institute of Finance and Statistics, University of Bonn, 53113 Bonn, Germany}
\thanks{Hausdorff Center for Mathematics, University of Bonn, 53113 Bonn, Germany},
Liebl D.\samethanks[2] \samethanks[3] \samethanks[4],
Mensinger T.\samethanks[3],
Gualtieri J.\thanks{Financial Services Transfer Pricing, Ernst \& Young Tax Co., Tokyo, Japan. The opinions expressed in this article belong to the author(s) and do not represent the views of Ernst \& Young network member firms.},
and
Sickles R.~C.\thanks{Department of Economics, Rice University, Houston, Texas 77005, USA}
}

\renewcommand\Authands{ and }


\date{}

\maketitle

\begin{abstract}
While a substantial literature on structural break change point analysis exists for univariate time series, research on large panel data models has not been as extensive.  In this paper, a novel method for estimating panel models with multiple structural changes is proposed. The breaks are allowed to occur at unknown points in time and may affect the multivariate slope parameters individually.  Our method adapts Haar wavelets to the structure of the observed variables in order to detect the change points of the parameters consistently.  We also develop methods to address endogenous regressors within our modeling framework.  The asymptotic property of our estimator is established.  In our application, we examine the impact of algorithmic trading on standard measures of market quality such as liquidity and volatility over a time period that covers the financial meltdown that began in 2007.  We are able to detect jumps in regression slope parameters automatically without using ad-hoc subsample selection criteria. 

\end{abstract}

\newtheorem{rmk}{Remark}\newtheorem{theorem}{Theorem}
\newtheorem{lemma}{Lemma} \newtheorem{proposition}{Proposition}
\newtheorem{corollary}{Corollary} \newtheorem{pprop}{Proof}

\newcommand{\quotes}[1]{``#1''}

\vspace{1 cm}

\null \vfil

\clearpage
\pagenumbering{arabic} \setcounter{page}{1}

\section{Introduction}
Panel datasets with large cross-sectional dimensions and large numbers of time observations are becoming increasingly available due to the impressive progress of information technology. Parallel progress has also occurred in the econometric literature, by the development of new methods and techniques for analyzing large panels. An important issue that has been addressed in several contexts is the risk of neglecting structural breaks in the data generating process, especially when the observation period is large.

Our paper contributes to this literature by providing a general framework to estimate panel models with multiple structural changes that occur at unknown points in time and may affect the model parameters individually.  We also develop methods to address endogenous regressors within our modeling framework.

Given a dependent variable $Y_{it}$ observed for $i=1,\dots,n$ individuals at $t=1,\dots,T$ time points, we consider the model
\begin{equation}
Y_{it}=
\mu+
\sum_{p=1}^{P}\sum_{j=1}^{S_{p}+1}X_{it,p}\mathbf{I}\big(\tau_{j-1,p}<t\leq \tau_{j,p}\big)\beta_{\tau_{j},p}+
\alpha_{i}+
\theta_{t}+
\varepsilon_{it}, \label{intromodel}%
\end{equation}
where $\mathbf{I}(.)$ is the indicator function, $X_{it,p}$, $p=1,\dots,P$, are explanatory variables, $\alpha_{i}$ is an individual specific effect, $\theta_{t}$ is a common time parameter, and $\varepsilon_{it}$ is an unobserved idiosyncratic term that may be correlated with one or more explanatory variables. For each variable $X_{it,p}$, $p=1,\ldots,P$, the corresponding slope parameter is piecewise constant with an unknown set of jump points
$\{\tau_{0,p},\tau_{1,p},\allowbreak \ldots,\allowbreak \tau_{S_{p}+1,p}|\  \allowbreak \tau_{0,p}=0<\tau_{1,p}<\ldots<\tau_{S_{p}+1,p}=T\}\subseteq \{0,1,\ldots,T\}$  for unknown numbers $S_{p}\geq 1$.

Our main contributions are as following. To the best of our
knowledge, we are the first to allow for different break points in each of the
model parameters which is of great practical importance since structural breaks
do not necessarily affect all explanatory variables simultaneously. We do not
impose restrictive assumptions on the number, the location, and/or the aspect of
the breaks. Our
approach is very general and covers the important situations encountered in
panel data analysis; the method can be applied to panel data with large time
dimension $T$ and large cross-section dimension $n$ and allows for $T$ to be
very large compared to $n$. We also consider the classic case of panel data, in
which $T$ is fixed and only $n$ is large.  Moreover,  we also show that our
Structure Adapted  Wavelet (SAW) estimator achieves mean square consistency
even if the cross-section dimension $n$ is small and fixed. The latter property
is a unique theoretical property made possible by using a Haar wavelets basis
function approach. 
Our estimation procedure adapts Haar wavelets to the structure of the observed explanatory variables and allows us to detect the location of the break points consistently.  We propose a general setup allowing for endogenous models such as dynamic panel models and/or structural models with simultaneous panel equations.  Consistency under weak forms of dependency and heteroscedasticity in the idiosyncratic errors is established and the convergence rate of our slope estimator is derived.  To consistently detect the jump locations and test for the statistical significance of the breaks,  we propose post-wavelet procedures.  We prove that our final estimator of the model parameters has the same asymptotic distribution as the (infeasible) estimators that would be obtained if all jump locations were exactly known a priori and thus possess the oracle property.  Our simulations show that, in many configurations of the data, our
method performs very well even when the idiosyncratic errors are affected by
weak forms of serial-autocorrelation and/or heteroscedasticity. 
Moreover, In the online Appendix A of our supplementary materials \citet{Bada2018b}, we
additionally discuss that our method can be extended to cover the case of panel models with unobserved heterogeneous common factors in the spirit of \citet{Ahn2001}, \citet{Pesaran2006}, \citet{Bai2009a}, \citet{Kneip2009}, \citet{Ahn2006},
and \citet{Bada2013}.

One of the earliest contributions to the literature on testing for structural
breaks in panel data is the work of \citet{Han1989}.  The authors propose a
multivariate version of the cusum-test, which can be seen as a direct extension
of the univariate time series test proposed by \citet{Brown1975}.
\citet{Qu2007} extend the work of \citet{Bai2003a} and consider the problem of
estimating, computing, and testing multiple structural changes that occur at
unknown dates in linear multivariate regression models. They propose a
quasi-maximum likelihood method and likelihood ratio-type statistics based on
Gaussian errors.  The method, however, requires that the number of equations is
fixed and does not consider the case of large panel models with unobserved
effects and possible endogenous regressors.  Based on the work of
\citet{Andrews1993}, \citet{DeWachter2012a} propose a break testing procedure
for dynamic panel data models with exogenous or pre-determined regressors when
$n$ is large and $T$ is fixed.  The method can be used to test for the presence
of a structural break in the slope parameters and/or in the unobserved fixed
effects.  However, their assumptions allow only for the presence of a single
break. \citet{Bai2010} proposes a framework to estimate the break in means and
variances. He also considers the case of one break and establishes consistency
for both large and fixed $T$. \citet{Kim2014} extends the work of
\citet{Bai2010} to allow for the presence of unobserved common factors in the
model. \citet{Pauwels2012} analyze the cases of a known and an unknown break
date and propose a Chow-type test allowing for the break to affect some, but not
all, cross-section units. Although the method concerns the one-break case, it
requires intensive computation to select the most likely individual breaks from
all possible sub-intervals when the break date is unknown.  \citet{Qian2014}
propose a three step procedure. In a first step they use a \quotes{na\"{\i}ve}
estimator of the slope parameter based on fits over the $n$ individuals for
every single time point $t$. This is used to construct appropriate weights for
the adaptive lasso procedure used in the second step. In a third step they then
propose a post lasso procedure. An important point is that their theory requires
that the preliminary na\"{\i}ve estimators of the slope parameters are
consistent (otherwise the weights in the adaptive lasso procedure may be
completely irregular). This of course will only work for large $n$. Their method
does not consider cases in which $n$ is small and $T$ is large.  Note that the
structure of panel data and the fact that the cross-section dimension $n$ may be
arbitrarily large (or small) makes it also difficult to justify a simple
reliance on methods developed in the literature on time series analysis. Time
series methods for break point detection are typically suffering from boundary
problems which can be mitigated using a panel-data approach.  Another problem when applying time series approaches to panel data is due to the typical individual effects structure. Error terms for the same individual are highly correlated and do not follow the usual assumptions made in the time series context.  A transformed model naively based on differences deals with the individual effects but does not fit either since then the jumping parameters $ \beta_{\tau_j,p} $ and $\beta_{\tau_j-1, p}$, $p = 1, \allowbreak, \ldots, \allowbreak P$, simultaneously occur in the equation.

Although our approach has some similarity to the likelihood based approach of
\citet{Li2015}, it is not clear how their approach could be implemented in a
panel data context. That appears not to be the case with the approach of
\citet{Bai1998, Bai2003a}, which could in principle also be used for panel data.
Indeed Bai and Perron are instructive in that they can help to motivate our
basic ideas. For a given number $S$ of breakpoints, the methods of Bai and
Perron are based on comparing the local fits of each possible combination of
$S+1$ subintervals. Although the method in \citet{Bai1998} is computationally
infeasible, the dynamic programming approach in \citet{Bai2003a} requires
$O(T^{2})$ fits of different subintervals. Our approach is also based on local fitting of
subintervals, using Haar wavelet ideas, which only needs to determine local
fits for $T$ subintervals. Another important advantage of our method over the methods of \citet{Bai1998, Bai2003a}, \citet{Qu2007}, and \citet{Qian2014} is that the elements of the
slope parameter vector are not forced to jump simultaneously.  Our method allows
for the total number of $S_{1}+\dots+S_{P}$ jumps at unknown jump locations
$\tau_{1,p}, \allowbreak \ldots, \allowbreak  \tau_{S_p,p}$ that are estimated
and identified individually for each regressor $p=1, \allowbreak \ldots,
\allowbreak P$.  Because the method of \citet{Bai1998, Bai2003a} is based on
optimizing the objective function over all possible sub-interval combinations,
allowing for the $P$ many slope parameters to jump individually will
dramatically increase the number of fits in their algorithm (exponentially on
the number of regressors). The method of \citet{Qian2014} also does not consider
the case of different structural breaks across the regressors and is also not
easily generalizable in this direction.  Individual jump locations for each
explanatory variable would generally require individual lasso-penalty parameters
for each of the $P$ many regressors.  This, however, would result then in a very
complex optimization problem for finding reasonable values for the lasso penalty
parameters.  Moreover, it is not clear how the proposed group-lasso will perform
in the case of correlated regressors.

Our empirical vehicle for highlighting this new methodology addresses the stability of the relationship between Algorithmic Trading (AT) and Market Quality (MQ). We propose to automatically detect jumps in regression slope parameters to examine the effect of algorithmic trading on market quality in different market situations. We find evidence that the relationship between AT and MQ was disrupted between $2007$ and $2008$, allowing firms that engage in AT to trade on variations in the relationship between liquidity and conventional measures of market quality.  This is a one rationale for the increased interest by regulators in the US and Europe who have proposed new market rules to monitor the positions taken by AT firms; see \citet{conghft}. The considered period coincides with the beginning of the subprime crisis in the US market and the bankruptcy of the big financial services firm Lehman Brothers and our findings have important implications for proponents and critics of high-frequency trading.

The remainder of the paper is organized as follows.  In Section
\ref{UnivPanMod}, we consider panel models with unobserved effects and multiple
jumping slope parameters, present our model assumptions, and derive the main
asymptotic results. Section \ref{postSAWproced} proposes a post-wavelet
procedure to estimate the jump locations, derives the asymptotic distribution of
the final estimator, and describes selective testing procedures. Section \ref{sumulations} presents the simulation results of our Monte Carlo experiments. Section \ref{application} focuses on the empirical application.  The conclusion follows in Section \ref{conclus}. The mathematical proofs are
collected in the online appendix of supplementary materials.

\section{Setup and Estimation Procedure} \label{UnivPanMod}
\subsection{Model}

Collecting the slope parameters in a $P\times 1$ time-varying vector, we can rewrite Model \eqref{intromodel} as
\begin{equation}
Y_{it}=\mu+X_{it}^{^{\prime}}\beta_{t}+\alpha_{i}+\theta_{t}+e_{it},
\label{multimodel}%
\end{equation}
where $X_{it}=(X_{1,it},\ldots,X_{P,it})^{^{\prime}}$ is the $P\times 1$ vector of regressors, $\beta_{t}=(\beta_{t,1}\ldots,\beta_{t,P})^{^{\prime}}$
is a unknown $P\times 1$ vector of slope parameters, and for each
$\beta_{t,p}$, $p=1,\ldots,P$, we have
\begin{equation}
\beta_{t,p}=\sum_{j=1}^{S_{p}+1}\mathbf{I}\big(\tau_{j-1,p}<t\leq \tau
_{j,p}\big)\beta_{\tau_{j},p},\label{kvbeta}%
\end{equation}
with $\tau_{0,p}=0$ and $\tau_{S_p+1,p}=T$ for all $p=1,\dots,P$. Equation \eqref{kvbeta} allows for each slope parameter $\beta_{t,p}$, $p=1,\ldots,P$, to change at $p$-specific unknown break points, $\tau_{1,p},\dots,\tau_{S_p+1,p}$.

Even in the absence of structural breaks, the uniqueness of $\mu$, $\alpha_{i}$ and $\theta_{t}$ requires some identification conditions -- we impose the commonly used restrictions:
\begin{equation*}%
\begin{array}
[c]{llcl}%
$C.1:$ & \sum_{i=1}^{n} \alpha_{i} & = & 0 \; \text{ and }\\[1ex]
$C.2:$ & \sum_{t=1}^{T} \theta_{t} & = & 0.
\end{array}
\end{equation*}
Note that the choice of C.1 and C.2 becomes irrelevant when the focus lies upon
estimating the slope parameters $\beta_{t,p}$, $p=1,\ldots,P$, but
  they are necessary to identify all model components in model
  \eqref{multimodel}, given a consistent estimate of the slope parameters
  $\beta_t$, by applying the usual within-, between-, and overall-average transformations.

In order to cover the case of dynamic models with both small and large $T$, we conventionally start with differencing the model to eliminate the individual effects and assume the existence of appropriate instruments.  Taking first order differences on the left and the right hand side of \eqref{multimodel} yields
\begin{equation}
\label{diffeq}\Delta Y_{it} = X_{it}^{^{\prime}}\beta_{t} - X_{i,t-1}%
^{^{\prime}}\beta_{t-1} + \Delta \theta_{t} + \Delta e_{it},
\end{equation}
for $i=1,\ldots,n$ and $t=2,\ldots,T$, where $\Delta$ denotes the first order difference operator.

The unobserved time-specific effect $\Delta \theta_{t} = \theta_{t} - \theta_{t-1}$
  can be eliminated by using a classic between transformation on the model, i.e., transforming
  $\Delta Y_{it}$ to $\Delta \dot{Y_{it}} = \Delta Y_{it} - \allowbreak n^{-1}
  \sum_{i=1}^{n} \Delta Y_{it}$. In this case $\Delta \theta_{t}$ is not part
  of the parameter vector $\gamma_t$ in \eqref{ldiffeq}, but otherwise our theoretical results
  remain valid.  Alternatively, we can associate $\Delta \theta_{t}$ with an additional unit
  regressor in the model and estimate it jointly with $\beta_{t+1}$ as a potential
  jumping parameter. Indeed, allowing for $\Delta \theta_{t}$ to be piecewise
  constant over time can be very useful for interpretation and means that the
  original time effect $\theta_{t} $ is a piecewise linear trend function
  possibly changing at every time point $t$.

Letting $\underline{X}_{it}= (X_{it}^{^{\prime}}, - X_{i,t-1}^{^{\prime}}, 1)^{^{\prime}} $ and $\gamma_{t} = (\beta_{t}^{^{\prime}}, \beta_{t-1}^{^{\prime}}, \Delta \theta_{t})^{^{\prime}}$ be $\underline P \times 1$ vectors with $\underline P = 2P + 1$ allows us to write Model \eqref{diffeq} as
\begin{equation}
\label{ldiffeq}%
\begin{array}
[c]{lcl}%
\Delta Y_{it} & = & (X_{it}^{^{\prime}}, - X_{i,t-1}^{^{\prime}}, 1) \left(
\begin{array}
[c]{c}%
\beta_{t}\\
\beta_{t-1}\\
\Delta \theta_{t}%
\end{array}
\right)  + \Delta e_{it},\\
& = & \underline{X}_{it}^{^{\prime}} \gamma_{t} + \Delta e_{it},
\end{array}
\end{equation}
for $i=1,\ldots,n$ and $t=2,\ldots,T$.

After taking differences, the time dimension consists of $T-1$ time periods.
For technical reasons, we need to assume that $T-1$ is \textit{dyadic}, i.e.,
$T-1=2^{L-1}$ for some positive integer $L\geq2$.  This is a technical
assumption that is only required for constructing the wavelet basis introduced
in the next section.  In practice, such an assumption does not impose any
restriction, since one can always  replicate the data by reflecting the
observations at the boundaries until getting the desired dimension.  If, for
instance, $T-1=125$, we can extend the too short sample $(\Delta Y_{i2},\allowbreak
\underline{X}_{i2}),\ldots,(\Delta Y_{iT},\allowbreak\underline{X}_{iT})$ by
adding the three last observations in reversed order; i.e.~by adding $(\Delta
Y_{i,T+1},\allowbreak \underline{X}_{i,T+1}):=(\Delta
Y_{iT},\allowbreak \underline{X}_{iT})$, $(\Delta
Y_{i,T+2},\allowbreak \underline{X}_{i,T+2}):=(\Delta Y_{i,T-1},\allowbreak
\underline{X}_{i,T-1})$, and $(\Delta
Y_{i,T+3},\allowbreak \underline{X}_{i,T+3}):=(\Delta Y_{i,T-2},\allowbreak
\underline{X}_{i,T-2})$ to achieve a dyadic time dimension $2^{7}=128$. That is,
generally, for any non-diadic $T-1$, one would prolong the sample by $(\Delta
Y_{i,T+1},\allowbreak \underline{X}_{i,T+1}):=(\Delta
Y_{iT},\allowbreak \underline{X}_{iT})$, $\dots$, $(\Delta
Y_{i,T+m},\allowbreak \underline{X}_{i,T+m}):=(\Delta Y_{i,T-(m-1)},\allowbreak
\underline{X}_{i,T-(m-1)})$, where $m=2^{g}-(T-1)$ with $g$ being the smallest
integer such that $2^{g}>(T-1)$.\\
Of course, any break-point detected in the synthetically prolonged time-range
from $T+1$ to $T+m$ would be meaningless and, therefore, simply removed from the
results.  Given the detected jump-locations, our post-SAW estimator described in
Section \ref{postSAWproced} uses then again the original, usually non-diadic, time dimension $T-1$.

\subsection{Some Fundamental Concepts of Wavelet Transformations}

The basic idea of our estimation approach consists of using a structurally adapted  Haar wavelet expansion of $\gamma_{t}$ to control for its piecewise changing character.  In order to motivate and explain our estimation method, we first need to introduce some important concepts and notations.  A general introduction to wavelet theory can be found, for instance, in \citet{Ruch2009}.

Let's start by considering a univariate step function $g:[1,T^*]\to\mathbb{R}$
with $T^*=2^{L-1}$.\footnote{The non-parametric time-series literature typically
  considers the normalized case for smoothing in the
  time-domain, i.e.~where $g:t/T^\ast\in[0,1]\mapsto g(t/T^\ast)\in
  \mathbb{R}$ \citep[see, for instance,][Ch.~6]{fan2008nonlinear}. This
  difference in normalization needs to be taken into account, but is effectively
  without loss of generality.} 
As in the case of $\gamma_t$, we will usually write $g_t\equiv g(t)$ since $g(\cdot)$ is only evaluated at the countable time points $t=1,\dots,T^*$. We use the diadic number $T^*$ for non-differenced data to distinguish it from the above introduced diadic number $T-1$ for differenced data.  Technically, the discrete wavelet expansion is much like the discrete Fourier transformation, which provides a decomposition of $g_t$ in terms of $T^*$ orthonormal basis functions of $\mathbb{R}^{T^*}$. The crucial difference, however, consists in the fact that wavelet bases are functions with local support.  The basis functions are defined as
\begin{align}
\psi_{l,k}(t) = \sqrt{2^{l-2}}\big(I_{l,2k-1}(t)- I_{l,2k}(t)\big)\label{varphi}
\end{align}
with $l=2,\dots,L$, $L\geq 2$, $k=1,\dots,K_l$, and $K_l=2^{2l-2}$,  where ${I}_{l,m}(t)$ is the indicator function which equals one if $t=2^{L-l}(m-1)+1,\ldots,2^{L-l}m$ and zero otherwise. Therefore, only $2^{L-l+1}$ functional values of $\psi_{l,k}(t)$ are nonzero.  The simplest version of a so-called Haar wavelet decomposition can then be written in the following form
\begin{equation}
g_{t}=c_1+\sum_{l=2}^{L}\sum_{k=1}^{K_{l}}\psi_{l,k}(t)c_{l,k}\quad\text{with}\quad t=1,\ldots,T^*.\label{betaweorigine}
\end{equation}
For any vector $g=(g_1,\dots,g_{T^*})'\in \mathbb{R}^{T^*}$ there exist coefficients $c_1$, $c_{2,1},\dots,c_{L,K_L}$ such that $g$ can be represented by \eqref{betaweorigine}, where the vectors $(1,\dots,1)'$ and $\boldsymbol{\psi}_{l,k}=(\psi_{l,k}(1),\dots,\psi_{l,k}(T^*))'$ establish an orthogonal basis of $\mathbb{R}^{T^*}$.

Wavelets are widely used in nonparametric regression analysis with the aim to estimate the regression function $g_t$ from noisy observations $z_t$ with
\begin{equation}
z_{t}=g_t +\epsilon_t\quad\text{with}\quad t=1,\ldots,T^*, \label{nonparregsimp}
\end{equation}
where $\epsilon_t$ are i.i.d.~zero mean error terms with finite variance
$\sigma^2$. Haar wavelets are particularly useful in situations where $g_t$ is a
piecewise constant function. Wavelet theory for standard nonparametric regression problems is not directly usable in our more complicated context.  However, we will show that the following basic properties and procedures can be generalized to our situation:
\begin{itemize}
\item[1)] Haar wavelets allow for a {\em sparse} representation of piecewise constant functions. If $g_t$ has exactly $S\ll T^*$ jumps, then \eqref{betaweorigine} holds with at most $(S+1)L$ non-zero coefficients $c_{l,k}$ with $l=2,\dots,T^*$ and $k=1,\dots,K_l$.
\item[2)] Orthogonality of the wavelet basis implies that least squares estimates of the coefficients in model \eqref{betaweorigine} are obtained independent of each other, $\tilde c_1 = \frac{1}{T^*} \sum_{t=1}^{T^*} z_{t}$ and $\tilde c_{l,k}=\frac{1}{T^*} \sum_{t=1}^{T^*}\psi_{l,k}(t) z_t$ with $l=2,\dots,T^*$ and $k=1,\dots,K_l$.
\item[3)] Wavelet thresholding (see, for instance, \citet{Donoho1994a}): If $\epsilon_t$ are normally distributed, then with probability tending to $1$ we have $|\tilde c_{l,k} -c_{l,k}|\leq A \sqrt{\log T^*/T^*}$ for some $A>\sqrt{2}$. Denoising by thresholding then means to estimate $c_{l,k}$ by $\hat c_{l,k}:=\tilde c_{l,k}$ if $\tilde c_{l,k}>A \sqrt{\log T^*/T^*}$ and by $\hat c_{l,k}:=0$ otherwise.  Then $\hat g_t :=\hat c_1+\sum_{l=2}^{L}\sum_{k=1}^{K_{l}}\psi_{l,k}(t)\hat c_{l,k}$.  For a piecewise constant function with $S\ll T^*$ jumps  (see Theorem \ref{theobetaSAW}), one gets the following rate of convergence: $(T^*)^{-1} \sum_{t=1}^{T^*} (g_t-\hat g_t)^2 =O_P((S+1)L\log T^*/T^*)$ as $T^*\rightarrow\infty$.
\end{itemize}

Our setup implies that for any time period $t$ we have to deal with a multivariate, namely, $\underline P$-dimensional vector $\gamma_t$ for some $\underline P>1$ (see Model \eqref{ldiffeq}).  A basic insight now is that a sparse representation of piecewise constant trajectories can be obtained when modifying \eqref{betaweorigine} by introducing sets of multivariate wavelet basis functions.  Let $A_{1,1}$ and $A_{l,k}$ with $l=2,\dots,L$ and $k=1,\dots,K_l$, denote arbitrary, non-singular $\underline P\times \underline P$ matrices, and set
\begin{equation}
\label{Wlk11}W_{l,k}(t) = \left \{
\begin{array}[c]{lll}%
A_{1,1}  & \text{if } & l= 1\\
A_{l,2k-1}H_{l, 2k-1}(t) - A_{l,2k}H_{l,2k}(t) & \text{if} & l > 1
\end{array}
\right.
\end{equation}
with
\[
H_{l, m}(t) = \sqrt{2^{l-2}} I_{l,m}(t-1),
\]
where $I_{l,m}(.)$ is defined as above.  The multivariate basis functions $W_{l,k}(t)\in\mathbb{R}^{\underline{P}\times\underline{P}}$ allow a multivariate wavelet expansion of $\gamma_{t}\in\mathbb{R}^{\underline P}$ such that
\begin{equation}
\label{undbetat}\gamma_{t} = \sum_{l = 1}^{L}\sum_{k = 1}^{K_{l}} W_{lk}(t)
\underline{b}_{l,k}\quad\text{for}\quad t=2,\ldots,T,
\end{equation}
where $L\geq 2$ and $K_l=2^{2l-2}$, with $\underline{b}_{l,k}\in\mathbb{R}^{\underline{P}\times 1}$.

Similar to Haar wavelets in a univariate setup, \eqref{undbetat} allows for a sparse representation of the piecewise constant parameter function $\gamma_t$:
\begin{proposition}\label{cor1}
Consider a sequence $\gamma_2,\dots,\gamma_{T}$ of  $\underline P$-dimensional vectors,  $\underline P\geq 1$. Suppose that for some $S<T-1$ there exist time points $\{\tau_1,\dots,\tau_{S}\}\subset \{2,\dots,T\}$ such that $\gamma_{t+1} \neq \gamma_{t}$ whenever $t\in \{\tau_1,\dots,\tau_{S}\}$ and $\gamma_{t+1} = \gamma_{t}$ otherwise. Then for any choice of invertible matrices $A_{l,k}$ the resulting expansion \eqref{undbetat}  holds with at most $(S+1)L$ coefficient vectors $\underline{b}_{l,k}$ satisfying $\underline{b}_{l,k}\neq 0$.
\end{proposition}

Proposition \ref{cor1} highlights the particular advantage of a multivariate
Haar wavelets basis in the special case of a sparse number of $S$ jumps. Our theoretical
results, however, allow also the challenging case, where the number of
paramter jumps,
$\sum_{p=1}^{\underline{P}}S_p$, 
diverges together with the number of time points $T$; see result \textit{(ii)}
of Theorem \ref{theo1}.

Note that expansion \eqref{undbetat} does not uniquely define the invertible matrices $A_{l,k}$.  Our estimation procedure defined below will rely on a particular choice of these matrices which adapts the typical wavelet orthogonality conditions (see point 2 above) for our panel data setup. This then allows a statistically efficient estimation procedure based on thresholding.  As this is one of our central contributions, we refer to our estimators as Structure Adapted Wavelet (SAW) estimators.

\subsection[]{Structure Adapted Wavelet Estimation of $\gamma_t$}

Following the above arguments our approach is based on representing $\gamma_t$ in \eqref{diffeq} by the multivariate wavelet decomposition \eqref{undbetat} relying on  suitable matrices $A_{lk}$ in \eqref{Wlk11}.  Our procedure relies on estimating the multivariate basis coefficients $\underline b_{l,k}$ which then lead to a corresponding estimate of $\gamma_t$.

Throughout, we assume the existence of a $\underline P \times1$ vector of
instruments $\underline Z_{it}$ which is correlated with $\underline X_{it}$ and
fulfills the exclusion restriction $E(\underline Z_{it}\Delta e_{it}) = 0$ for
all $i$ and $t$.  The unit regressor associated with $\Delta \theta_{t}$ and the
remaining exogenous regressors (if they exist) can, of course, be instrumented
by themselves.

\begin{rmk}
  In the over-identified case, where there are more than $\underline P$ many
    instruments, one can use a classic two-stage least squares approach. In this case,
    each element in $\underline Z_{it}\in\mathbb{R}^{\underline P \times1}$ denotes
    the fitted value from regressing each of the $\underline P$ many elements in $\underline X_{it}$ on all instruments.
\end{rmk}

In view of \eqref{undbetat}  let $\underline{\mathcal{Z}}_{l,k,it}^{^{\prime}} = \underline Z_{it}%
^{^{\prime}} W_{l,k}(t)$ and $\underline{\mathcal{X}}_{l,k, it}^{^{\prime}} = \underline
X_{it}^{^{\prime}} W_{lk}(t)$. We then have $\underline{X}_{it}^{^{\prime}} \gamma_{t}=\sum_{l=1}^{L}\sum_{k=1}^{K_{l}}\underline{\mathcal{X}}_{l,k, it}^{^{\prime}} \underline b_{l,k}$ and thus $\underline{\mathcal{X}}_{l,k, it}$ constitute modified regressors to be used for estimating the multivariate basis coefficients $\underline b_{l,k}$.  Correspondingly, $\underline{\mathcal{Z}}_{l,k, it}$ represent the modified instruments.  Note that since $E(\underline Z_{it}\Delta e_{it}) = \allowbreak0$ for all $i$ and $t$, we can infer that $E(\underline{\mathcal{Z}}_{l,k, it}\Delta e_{it}) = \allowbreak0$ for all $l$ and $k$.  Relying on $\underline{\mathcal{Z}}_{l,k, it}$, $\underline{\mathcal{X}}_{l,k, it}$,  the
required theoretical moment condition for estimating $\underline b=(\underline
b_{1,1},\dots,\underline b_{L,K_L})$ is 
\begin{equation}
\label{MM}E\left(\sum_{l=1}^{L}\sum_{k=1}^{K_{l}}\underline{\mathcal{Z}}_{l,k, it}(\Delta Y_{it} - \underline{\mathcal{X}}_{l,k, it}^{^{\prime}} \underline b_{l,k})
\right) = 0.
\end{equation}
The IV estimator, $\underline{\tilde{b}}_{l,k}$, of $\underline b_{l,k}$ is the solution to the empirical moment condition
\begin{equation}
\label{multempIV}\frac{1}{n(T-1)}\sum_{i=1}^{n} \sum_{t=2}^{T} \sum_{l=1}%
^{L}\sum_{k=1}^{K_{l}}\big( \underline{\mathcal{Z}}_{l,k,it}(\Delta Y_{it} -
\underline{\mathcal{X}}_{l,k,it}^{^{\prime}} \underline{\tilde{b}}_{l,k} )
\big) = 0.
\end{equation}

Under general assumptions, it is possible to state the consistency of the IV
estimator, $\underline{\tilde{b}}_{l,k}$, of $\underline b$ for all possible
matrices $A_{lk}$ in \eqref{Wlk11}.  However, computational efficiency can be
enhanced by sophisticated choices of $A_{lk}$ satisfying orthonormality
conditions which imply that each basis coefficient vector
$\underline{\tilde{b}}_{l,k}$ can be estimated separately from all other basis
coefficient vectors $\underline{\tilde{b}}_{l',k'}$ with $(l,k)\neq (l',k')$.  A
major advantage of this orthogonalization step consists in the fact that the
zero coefficients can be identified by a simple thresholding procedure.  By the
empirical moment condition in \eqref{multempIV} the relevant, data adaptive,
multivariate orthonormality conditions are given by
\begin{itemize}
\item[\textit{(A):}] $\displaystyle \frac{1}{n(T-1)} \sum_{i=1}^{n} \sum_{t=2}^{T}
\underline{\mathcal{Z}}_{l,k,it} \underline{\mathcal{X}}_{l^{^{\prime}%
  },k^{^{\prime}},it}^{^{\prime}}= I_{\underline P\times \underline P} $ if
$(l,k) = (l^{^{\prime}},k^{^{\prime}})$ and

\item[\textit{(B):}] $\displaystyle \frac{1}{n(T-1)} \sum_{i=1}^{n} \sum_{t=2}^{T}
\underline{\mathcal{Z}}_{l,k,it} \underline{\mathcal{X}}_{l^{^{\prime}%
},k^{^{\prime}},it}^{^{\prime}} = \mathbf{0}_{\underline P\times \underline P}
$ for all $(l,k) \neq(l^{^{\prime}},k^{^{\prime}})$,
\end{itemize}
where  $I_{\underline P\times \underline P} $ is the
$\underline P\times \underline P$ identity matrix and $\mathbf{0}_{\underline
P\times \underline P}$ is a $\underline P\times \underline P$ matrix of zeros.
Conditions (A) and (B) are multivariate orthonormality
  conditions referring to inner products of two $(T-1)\underline P\times \underline
  P$-matrices
  $$\displaystyle 
  m_{(l,k),(l',k')}=\left(\sqrt{\frac{\sum_{i=1}^{n}
        \underline{\mathcal{Z}}_{l,k,i2}\underline{\mathcal{X}}_{l',k',i2}}{n(T-1)}},\dots,
    \sqrt{\frac{\sum_{i=1}^{n}
        \underline{\mathcal{Z}}_{l,k,iT}\underline{\mathcal{X}}_{l',k',iT}}{n(T-1)}}\right)'$$
  where $m'_{(l,k),(l',k')}m_{(l,k),(l',k')}=I_{\underline P\times \underline P}$ if
  $(l,k)=(l',k')$ and $m'_{(l,k),(l',k')}m_{(l,k),(l',k')}=\mathbf{0}_{\underline P\times \underline P}$ if
       $(l,k)\neq(l',k')$. The special case $\underline{P}=1$ leads to classic
       univariate orthonormality conditions between two $(T-1)$-vectors.

Some straightforward calculations now show that conditions \textit{(A)} and \textit{(B)} are fulfilled if the matrices $A_{lk}$ in \eqref{Wlk11} are determined as
\[%
\begin{array}
[c]{lcl}%
A_{1,1} & = & \underline{Q}_{1,1}^{- \frac{1}{2}},\\
A_{l, 2k-1} & = & \underline{Q}_{l,2k-1}^{- 1} \big(\underline{Q}_{l,2k-1}^{-
1} + \underline{Q}_{l,2k}^{- 1}\big)^{- \frac{1}{2}}, \text{ and }\\
A_{l, 2k} & = & \underline{Q}_{l,2k}^{- 1} \big( \underline{Q}_{l,2k-1}^{- 1}
+ \underline{Q}_{l,2k}^{- 1}\big)^{- \frac{1}{2}},
\end{array}
\]
where
\[%
\begin{array}[c]{lcl}%
\underline{Q}_{1,1} & = & \frac{1}{n(T-1)}\sum_{i=1}^{n} \sum_{t = 2}^{T}
\underline Z_{it} \underline X_{it}^{^{\prime}},\\
\underline{Q}_{l,2r} & = & \frac{1}{n(T-1)} \sum_{i=1}^{n} \sum_{t = 2}^{T}
                               \underline Z_{it} \underline X_{it}^{^{\prime}}
                             H_{l,2r}(t)^{2}, \text{ for }r=\{k,k-1\}.\\
\end{array}
\]
Solving \eqref{multempIV} for $\underline{\tilde{b}}_{l,k}$ under the
normalization conditions $(A)$ and $(B)$ leads to the IV estimator
\[
\underline{\tilde{{b}}}_{l,k} = \frac{1}{n(T-1)} \sum_{i=1}^{n} \sum_{t = 2}^{T}
\underline{\mathcal{Z}}_{l,k,it}\Delta Y_{it}
\]
or alternatively
\[
\tilde{{b}}_{l,k,p} = \frac{1}{n(T-1)} \sum_{i=1}^{n} \sum_{t = 2}^{T}
{\mathcal{Z}}_{l,k, it, p}\Delta Y_{it},
\]
where $\tilde{{b}}_{l,k,p}$ and ${\mathcal{Z}}_{l,k, it, p}$ are the $p$th elements of $\tilde{\underline{b}}_{l,k}$ and $\underline{\mathcal{Z}}_{l,k,it}$, respectively.

Making use of the constructed orthonormality, we can now directly apply the universal thresholding scheme of \citet{Donoho1994a} to eliminate the insignificant coefficients. Let $\lambda_{nT}$ be a predetermined threshold and $\hat{{b}}_{l,k,q}$ the estimator of the wavelet coefficients after shrinkage, i.e.,
\begin{equation}
\hat{{b}}_{l,k,q} = \left \{
\begin{array}
[c]{lll}%
\tilde{b}_{l,k,q} & \text{ if } & |\tilde{b}_{l,k,q}| > \lambda_{nT} \;
\text{ and }\\
0 & \text{ else. } &
\end{array}
\right.
\end{equation}
Our Structure Adapted Wavelet (SAW) estimators of the parameters $\gamma_{t,p}$, $p=1,\ldots,\underline P$, composing the vector $\gamma_{t}$ are then  obtained by
\begin{equation}\label{eq:SAW}
\hat{\gamma}_{t,p} = \sum_{l=1}^{L}\sum_{k=1}^{K_{l}}\sum_{q=1}^{\underline P}W_{lk,p,q}(t)\hat{{b}}_{l,k,q},
\end{equation}
where $W_{lk,p,q}$ is the $(p,q)$-element of the basis matrix $W_{lk}(t)$ with $p,q=1,\ldots,\underline{P}$.  The threshold parameter $\lambda_{nT}$ depends on the sample dimensions $n$ and $T$ and converges to zero as $nT\to\infty$; see our asymptotic results in the next section.

Recall that by construction $\gamma_{t,p}=\beta_{t,p}$, $\gamma_{t,p+P}=\beta_{t-1,p}$ for $p=1,\ldots,P$ and $\gamma_{t,p}=\Delta \theta_{t}$ for $p=\underline{P}$.  Therefore, a first step estimator of $\beta_{t,p}$ for $t=2,\ldots,T$ can be obtained by $\hat{\gamma}_{t,p}$.  Another natural estimator of $\beta_{t,p}$ for $t=1,\ldots,T-1$ is $\hat{\gamma}_{t+1,p+P}$. A first step estimator of $\Delta \theta_{t}$ can be obtained by $\hat{\gamma}_{t,\underline{P}}$.  In the next section, We show the uniform and the mean squared consistency of $\hat{\gamma}_{t,p}$ for all $p=1,\allowbreak \ldots,\allowbreak \underline{P}$.  If, however, $n$ is very large, we propose to use this estimator only as a \emph{first step} estimator for estimating the jump time points $\tau_{1,p},\ldots,\tau_{S_{p},p}$, $p=1,\ldots,P$.  Once the jump time points are detected, we propose to perform a post-SAW estimation in order to achieve more efficient parameter estimates and permitting classical inferences; see Section \ref{postsawestimationvk}.

\begin{rmk}
The theoretical moment condition \eqref{MM} holds for any fixed $W_{lk}$, but it may fail to hold exactly for our
data dependent definition of the matrices $A_{lk}$. This is, however, of no importance since in any case
$\frac{1}{n(T-1)}\sum_{i=1}^{n} \sum_{t=2}^{T} \sum_{l=1}^{L}\sum_{k=1}^{K_{l}}\allowbreak
\underline{\mathcal{Z}}_{l,k,it}\Delta e_{it}$ can  be expected to be close to
zero. The reason is that this term consists of  partial sums  of the form
$A_{l,2k-1}\big(\frac{1}{n(T-1)}\sum_{i=1}^{n}\sum_{t=2}^{T} H_{l,
  2k-1}(t)\underline Z_{it} \Delta e_{it}\big)$ and
$A_{l,2k}\allowbreak\big(\frac{1}{n(T-1)}\allowbreak\sum_{i=1}^{n}\sum_{t=2}^{T}
H_{l, 2k}(t)\underline Z_{it} \Delta e_{it}\big)$. Since $H(\cdot)$ is a
deterministic function, the partial moment conditions
$E(\sum_{t=2}^{T}\allowbreak H_{l, 2k-1}(t)\underline Z_{it} \Delta e_{it})=0$ and $E(\sum_{t=2}^{T}\allowbreak H_{l, 2k}(t)\underline Z_{it} \Delta e_{it})=0$ remain valid and provide a sensible basis for the proofs of our theoretical results.
\end{rmk}

\begin{rmk}\label{rmkdummy}
Considering Model \eqref{ldiffeq}, one may be tempted to naively use the available cross-section data to construct estimates $\hat\gamma_t$ separately for each time point $t$.  However, the resulting estimators would be highly correlated and statistically inefficient. In particular, any differentiation between $\gamma_t-\gamma_{t-1}=0$ and $\gamma_t-\gamma_{t-1}\neq 0$ in the presence of
noise becomes extremely problematic. Our basis functions can be seen as a special design of dummy vectors that has many technical and computational advantages.  We show below that our structure adapted wavelet estimator achieves mean square consistency even if the cross-section dimension $n$ is fixed.  This is not possible when naively using  $T$  dummy vectors. We also want to emphasize that in order to detect all possible structural changes, we  only need $T$ fits, while, for instance, \citet{Bai2003a} need $O(T^2)$ subinterval fits to do this in a time series context.
\end{rmk}

\begin{rmk}
The first step estimation of Model \eqref{multimodel} is based on the SAW estimation of the transformed Model \eqref{diffeq}, which is performed without taking into account the fact that $\gamma_{t,p}$ and $\gamma_{t+1, p + P}$ are identical time series of parameters. Although the estimators  $\hat{\gamma}_{t,p}$ and $\hat{\gamma}_{t+1,p+P}$ are not restricted to be identical for $t=2, \allowbreak \ldots, \allowbreak T-1$, they present a beneficial property that can be exploited for consistently estimating the jump locations; see Section \ref{detlectjumploca}. The loss of efficiency due to the extended number of parameters can be redressed through a post-SAW estimation once the jump locations are consistently estimated; see Section \ref{postsawestimationvk}.
\end{rmk}

\subsection{Assumptions and Main Asymptotic Results}\label{Asympt}
Throughout, we use $E_{c}(.)$ to define the conditional expectation given $\{X_{it}\}_{i,t\in \mathbb{N}^{\ast2}}$ and $\{Z_{it}\}_{i,t\in\mathbb{N}^{\ast2}}$, where $\mathbb{N}^{\ast}=\mathbb{N}\setminus \{0\}$.  We denote by $M$ a finite positive constant, not dependent on $n$ and $T$. The operators $\overset{p}{\longrightarrow}$ and $\overset{d}{\longrightarrow}$ denote the convergence in probability and distribution. $O_{p}(.)$ and $o_{p}(.)$ are the usual Landau-symbols. The Frobenius norm of a $p\times k$ matrix $A$ is denoted by $||A||=[\operatorname{trace}(A^\prime A)]^{1/2}$, where $A^\prime$ denotes the transpose of $A$.  In the following we present the assumptions of our asymptotic analysis.

\paragraph*{\textbf{Assumption A} (data dimensions):}
\label{assumptionA}
\begin{enumerate}
\item[(i)] $T  = 2^{L-1} + 1$ for some natural number $L > 1$; the number of
regressors $P$ is fixed.
\item[(ii)] $n \to \infty$; $T$ is either fixed or passes to infinity simultaneously with $n$ such that $\log(T)^{\varepsilon}/n \to 0$ for all $\varepsilon > 0$.
\end{enumerate}

\paragraph*{\textbf{Assumption $\text{A}^\star$}:}\label{assumptionAprime}
As Assumption A, but with point (ii) substituted by
\begin{enumerate}
\item[(ii$^\star$)] $n$ is fixed, $T\to\infty$, and the total number of parameter jumps, $\sum_{p=1}^{\underline P}S_p$, is fixed.
\end{enumerate}

\paragraph*{\textbf{Assumption B} (regressors and instruments):}
\label{assumptionB}
\begin{enumerate}
\item[(i)] For all $i$ and $t$, $E(\underline Z_{it}\Delta e_{it}) = 0$.  For all
$l \in \{1, \ldots, L\}$ and $k \in \{1, \ldots, K_{l}\}$ with
$\underline{b}_{l,k}\neq 0$,
\[ \Vert \underline Q_{l,k}-\underline Q^{\circ}_{l,k}\Vert\rightarrow_P 0, \text{ where }
\underline Q_{l,k} = \frac{1}{n \cdot \sharp \{s | H_{l,k}(s) \neq0\}} \sum_{t
\in \{s | H_{l,k}(s) \neq0\}}\sum_{i=1}^{n} \underline Z_{it}\underline
X_{it}^{^{\prime}},
\]
where $\underline Q^{\circ}_{l,k}$ is a $\underline P \times \underline P$ full rank finite matrix with linearly independent eigenvectors, and where for all $n$ sufficiently large the smallest eigenvalue of $\underline Q_{l,k}$, $\lambda_{\underline P}(\underline Q_{l,k})$, is such that with probability one $\lambda_{\underline P}(\underline Q_{l,k})\geq c$ for some arbitrarily small constant $c>0$.

\item[(ii)] The moments $E||\underline Z_{it}||^{4}$ and $E||\underline X_{it}||^{4} $ are bounded uniformly in $i=1,\dots,n$ and $t=2,\dots,T$.

\end{enumerate}

\paragraph*{\textbf{Assumption C} (weak dependencies and exponential tails for
    $Z_{it}\Delta e_{it}$)}
\label{assumptionC}
\begin{enumerate}
\item[(i)] There exists an $M<\infty$ such that for $1\leq s< s'\leq T$ and any
$p=1,\dots,P$
\[
\sigma^2_{n,s,s^{^{\prime}};p}:=\frac{1}{n(s^{^{\prime}} - s +1)}E\left(\big[\sum_{i=1}^{n}\sum_{t=s+1}^{s^{^{\prime}}}   Z_{it,p}\Delta e_{it} \big]^2\right) \leq
M.
\]
Note that $\sqrt{n(s^{^{\prime}} - s +1)}\sigma^2_{n,s,s^{^{\prime}};p}=\text{Var}(\frac{1}{n(s^{^{\prime}} - s +1)}\sum_{i=1}^{n}\sum_{t=s+1}^{s^{^{\prime}}}   Z_{it,p}\Delta e_{it})$.
\item[(ii)] There exist some  an $0<\delta_0,\delta_1,\delta_2<\infty$ such that for $1\leq s< s'\leq T$ and any
$p=1,\dots,P$
\begin{equation}\label{berntype}
P\left(\frac{1}{\sqrt{n(s^{^{\prime}} - s +1)}}\left|\sum_{i=1}^{n}\sum_{t=s+1}^{s^{^{\prime}}}  Z_{it,p}\Delta e_{it}\right| \geq c\cdot \sigma_{n,s,s^{^{\prime}};p} \right) \leq \delta_0\exp(-\delta_1 c^{\delta_2})
\end{equation}
for any $ c > 0$.
\end{enumerate}

\paragraph*{Discussion of assumptions} Assumption A (i) specifies a dyadic condition on the intertemporal data size $T$. This is a technical assumption that is only required for constructing the dyadic wavelet basis functions. As mentioned earlier, in practice, one can always replicate the data by reflecting the observations at the boundaries to get the desired dimension. This strategy does not change the validity of our asymptotic results.
Assumption A (ii) allows for the time dimension $T$ to be very large compared to $n$.  Moreover, Assumption A (ii) considers also the classical case of panel data, in which $T$ is fixed and only $n \to \infty$.\\
Assumption B (i) requires that the probability limit of $\underline{Q}_{l,k}$
is a full rank finite matrix with linearly independent eigenvectors.  This is to
ensure that the matrix inverse via eigendecomposition exists also for
non-symmetric matrices, $\underline{Z}_{it}\underline{X}'_{it}$, involving
instrumental variables.  The assumption that the smallest eigenvalue of
$\underline Q_{l,k}$ is bounded away from zero almost surely represents a
regularity condition to ensure that also higher moments of the matrices $A_{lk}$
exist and is only sightly more restrictive than the classic no-multicollinearity
assumption. Under the asymptotic regime of Assumption A, Assumption B (i)
  allows for non-stationary time-series $(X_{it})_{t=1,\dots,T}$. Under the asymptotic regime of Assumption A$^\star$, Assumption B (i)
  requires stationary time-series $(X_{it})_{t=1,\dots,T}$ and a $l^\star>0$ such that $\underline{b}_{l,k}=0$ for all
  $l>l^\star$ and all $k\in\{1,\dots,K_l\}$ with $l^\star=O(\log(L))$,
  where $L\to\infty$ as $T=2^{L-1}+1\to\infty$. The latter condition on
  $l^\star$ assures that there are asymptotically infinitely many potential jump-locations. Assumption B (ii) specifies commonly used moment conditions to allow for some limiting terms to be $O_{p}(1)$ when using Chebyshev inequality. \\
The main contents of Assumption C are a) sufficiently weak dependencies between the variables, and b)
existence and regularity of higher moments of $Z_{it,p}\Delta e_{it}$. Suppose that $E(Z_{it,p}^2\Delta e_{it}^2)\leq M$ and $E(Z_{it,p}^k\Delta e_{it}^k)\leq \frac{1}{2} k!M B^k$ for all $k>3$ and some constant  $B>0$. If all $Z_{it,p}\Delta e_{it}$ are independent, condition (i) is obviously satisfied, and Bernstein's inequality implies that 
$P(\frac{1}{\sqrt{n(s^{^{\prime}} - s +1)}}|\sum_{i=1}^{n}\sum_{t=s+1}^{s^{^{\prime}}}  Z_{it,p}\Delta e_{it}| \geq c\,\sqrt{M}) \allowbreak\leq\allowbreak
 2\exp\big(-\frac{n(s^{^{\prime}} - s +1)Mc^2}{2(M n(s^{^{\prime}} - s
    +1)+c\sqrt{Mn(s^{^{\prime}} - s +1)B})}\big) \allowbreak
  \leq 2 \max\{ \exp\big(-\frac{c^2}{4}\big),
  \exp\big(-\frac{c\sqrt{Mn(s^{^{\prime}} - s +1)}}{4B}\big)\}.$
  For independent variables this inequality even implies that \eqref{berntype} holds with $\delta_2=2$ if the sample is large enough such that $c<\sqrt{Mn(s^{^{\prime}} - s +1)}/B$. In practice,
a main source of dependencies will consist in correlations between $Z_{it,p}^k\Delta e_{it}$ and $Z_{i,t+1,p}^k\Delta e_{i,t+1}$ at successive time points, and we want to note that there also exist Bernstein type inequalities for time series data (assuming Markov chain or martingale structure). In this paper we only additionally assume that existing dependencies are sufficiently weak such that (i) holds and such that tail probabilities still decay at some
exponential rate as required by (ii).

The following Lemma establishes the main asymptotic results for our structure
adapted wavelet estimators, $\tilde{b}_{l,k,q}$, of the basis coefficients
${b}_{l,k,q}$.

\begin{lemma}\label{theo1}
Suppose Assumptions A, B, and C hold, then
\begin{itemize}
\item [(i)] $\displaystyle \sup_{l, k, q}   \left|\tilde{b}_{l,k,q} - {b}_{l,k,q} \right| = O_p \big(\log(T-1)^{1/\delta_2}\sqrt{1/n(T-1)}\big)$.
\item [(ii)] Moreover, there exists a constant ${\cal M}<\infty$ such that
  $$\sup_{l, k, q}   \left|\tilde{b}_{l,k,q} - {b}_{l,k,q} \right| \leq {\cal M}\log(T-1)^{1/\delta_2}\sqrt{1/n(T-1)}$$
holds with a probability that converges to $1$ independently of $n$, as $T \to \infty$.
\end{itemize}
\end{lemma}

Theorem \ref{theobetaSAW} establishes the uniform and the mean square
consistency of our SAW estimators $\hat\gamma_{t,p}$.

\begin{theorem}[SAW estimator]\label{theobetaSAW}
Suppose Assumptions A, B, and C hold, then
\begin{itemize}
\item[(i)]  $\sup_{t\in \{1, \ldots, T-1\}} |\hat \gamma_{t,p} - \gamma_{t,p}| = o_p(1) $ for all $p=1,\ldots,\underline{P}$, if $\sqrt{T-1} \lambda_{nT} \to 0$, as $n,T \to \infty$ or $n \to \infty$ and $T$ is fixed.
\end{itemize}
Suppose Assumptions A or A$^\star$, B, and C hold, then
\begin{itemize}
\item[(ii)] $\frac{1}{T-1}\sum_{t = 1}^{T-1}||\hat{\gamma}_{t} - \gamma_{t}||^2
  = O_{p}\big(\frac{(\log(T-1)^{2/\delta_2}/n)^{\kappa}J^{*}}{(T-1)}\big)$, if
  $\sqrt{T-1}\lambda_{n,T} \sim(\log(T-1)^{2/\delta_2}/n)^{{\kappa}/2}$ for any
  $0<\kappa<1$,  where $J^{*} = \min \{ \allowbreak (\sum_{p=1}^{\underline
    P}S_p +1)\log(T-1), (T-1)\}$, and where $0<\delta_2<\infty$ is from
  Assumption C (ii).
\end{itemize}
\end{theorem}
Result \textit{(i)} of Theorem \ref{theobetaSAW} shows that uniform consistency is obtained when $n \to \infty$ and $T$ is fixed or $n,T\to \infty$ with $\log(T)/n \to0$. Result \textit{(ii)} considers the mean square consistency of the parameter estimator sequence $\hat{\gamma}_1,\dots,\hat{\gamma}_T$.  For the challenging extreme case where the number of jumps is proportional to $T$, as $T\to\infty$, result \textit{(ii)} of Theorem \ref{theobetaSAW} shows that the mean square error, $(T-1)^{-1}\sum_{t = 1}^{T-1}||\hat{\gamma}_{t} - \gamma_{t}||^2$, converges with the rate of $\big(\log (T)^{\frac{2}{\delta_2}}/n\big)^\kappa$.

Our estimation approach is, however, particularly powerful if $\gamma_t$ is piecewise constant and the number of jumps remains bounded as $T\rightarrow\infty$.  Accuracy of parameter estimates then improves with $n$ and $T$ (Assumption A) as well as with $T$ while $n$ is fixed (Assumption A$^*$). From Theorem \ref{theobetaSAW} \textit{(ii)}, we can infer that in this case the mean square error of our SAW estimates of $\gamma_{t,p}$, $t=1,\ldots,T$, converges at a rate of $(\log T)^{1+\frac{2}{\delta_2}\kappa}/(Tn^\kappa)$ and at a rate of $(\log T)^{1+\frac{2}{\delta_2}\kappa}/T$ for fixed $n$ respectively. By contrast, the mean square error for the dummy approach discussed in Remark \ref{rmkdummy} only converges at a rate of $1/n$.  The explanation for this effect lies in the special structure of wavelets. It is a basis expansion, where each basis function is local and describes the behavior in a specific subinterval. Only the first basis function is global. The corresponding parameter is fitted over all $n(T-1)$ observations. With the fitted first basis function we essentially quantify the best global fit with a constant $b_{1,1}$. At each resolution level $l=2,3,\ldots$ we can then analyze parameter changes in $2^{(l-1)} $ local subintervals, which correspond to the respective basis functions. Each of these subintervals contains approximately $n(T-1)/(2^{(l-1)}-1)$ observation points. This means that the \textquotedblleft relative error\textquotedblright of the coefficient estimates increases as $l$ increases. The coefficients of the lowest level basis functions are only determined from $2n$ observations and are the least accurate as they are influenced by noise.

Accuracy of our SAW estimate also depends on whether or not the jump locations are dyadic. As an illustration, consider the case of one jump at some location $\tau = \tau_{1, 1} = \cdots = \tau_{1, P}$.  If $\tau=(T-1)/2$, then only the global fit and another basis function is necessary, and the mean squared error of the parameter estimates is proportional to $1/(nT)$. The same is true if, for instance, $\tau=3(T-1)/4,$ in which case one additional resolution level is required, but the mean squared error will still be\ proportional to $1/(nT)$. Also, with probability tending to $1$ the estimated location will converge to the true location even for fixed $n$. For arbitrary (non-dyadic) locations of the jump points, the inaccurate highest resolution level also will play a role. In this case for fixed $n$ the true point might not be exactly identified even for large $T$. Nevertheless, the information acquired from higher levels may still allow us to say with high probability that the true point lies, e.g., between $(6/8)(T-1)$ and $(7/8)(T-1)$. Theorem \ref{theobetaSAW} implies that the distance between the true and estimated location will be at least of order
$O_P(\log (T)^{1/2(1+(2/\delta_2)^\kappa)}/\sqrt{Tn^\kappa})$, i.e,
$\sqrt{T-1}|\hat\tau/(T-1)-\tau/(T-1)|= O_P(\log (T)^{1/2(1+(2/\delta_2)^\kappa)} / \sqrt{Tn^\kappa})$.
For fixed $n$ and simultaneous parameter jumps,  \citet{Bai2003a} arrive at
similar theoretical results if $\delta_2 = 2$ and thus our approach may be seen
as a computationally more efficient variant when there are only $T$ subinterval
fits.

Asymptotically, any threshold that has the following order 

\begin{equation}
\label{threshgood}
\lambda_{nT} \sim \left(\frac{\log(T-1)^{2/\delta_2}}{ n}\right)^{{\kappa/2}}\frac{1}{\sqrt{(T-1)}},
\end{equation}
with a strictly positive $\kappa <  1$ ensures the asymptotic assertions of Theorem \ref{theobetaSAW}. If the random variables ${Z}_{it,p}\Delta e_{it}$ are independent or $n, T$ satisfy Assumption A instead of $A^*$, then $\delta_2$ can be set to $2$ and the finite sample amplitude as well as $\kappa$ can be set analogously to the threshold defined in (\ref{threshgood}). In the ambiguous case where $n$ is fixed and ${Z}_{it,p}\Delta e_{it}$ are not independent, one can replace $\log(T-1)^{2/\delta_2}$ with $\sqrt{(T-1)}$.  This affects, of course, the convergence rate of $\sum_{t = 1}^{T-1}||\hat{\gamma}_{t} - \gamma_{t}||^2/(T-1)$ but still assures the mean square consistency even when $n$ is fixed. An explicit function of $\lambda_{nT}$ for our final post-SAW estimator, discussed below, is presented in Section \ref{detlectjumploca}.

\begin{rmk}
Summing up, the hierarchic step function structure of the Haar wavelet basis
functions allows us to profit from the data dimensions $n$ and/or $T$ --
depending on the context. This allows us to mitigate many known issues in classic
time series change-point detection methods, such as, for instance, asymptotic boundary points $\tau^B=\tau^B_T$ where $\tau^B_T/(T-1)\to 0$ or $\tau^B_T/(T-1)\to 1$
as $T\to\infty$. While such asymptotic boundary points are not in the focus
of this paper, we nevertheless can comment on this problem. Asymptotic
boundary points, $\tau_T^B$, will always require the highest wavelets resolution level $l$
which implies that we need the asymptotic scenario of Assumption A, where
$n\to\infty$ and $T$ may be fixed or $T\to\infty$ sufficiently slow.

Thus Theorem \ref{theobetaSAW} \textit{(ii)} implies that the mean squared error
of our SAW estimator, $(T-1)^{-1}\sum_{t = 1}^{T-1}||\hat{\gamma}_{t} -
\gamma_{t}||^2$, considering asymptotic boundary bounds, $\tau_T^B$, converges
with the rate of $1/(n^\kappa)$ instead of $(\log
T)^{1+\frac{2}{\delta_2}\kappa}/(Tn^\kappa)$. However, the information acquired from higher levels of the hierarchic Haar wavelets
basis would still allow us to say with high probability that there is a
boundary point, $\tau^B_T$, e.g., in the initial interval $[1,(1/8)(T-1)]$ or
the last interval $[(7/8)(T-1),(T-1)]$. More generally any initial and final
interval $[1,(1/2^{(l-1)})(T-1)]$ and
$[((2^{(l-1)}-1)/2^{(l-1)})(T-1),(T-1)]$ for any fixed $l=2,3,\dots$ can be considered.
Using similar arguments as in the proof
of Theorem \ref{theobetaSAW} \textit{(ii)}, one can show that the accuracy of
such interval-based statements would again be of the order $(\log
T)^{1+\frac{2}{\delta_2}\kappa}/(Tn^\kappa)$ allowing to profit from both data
dimensions $n$ and $T$.  
\end{rmk}

\begin{rmk}
So far, we have considered the estimation of jumping slope parameters in the case of stochastically bounded regressors; see Assumption B. Our method should also be appropriate for panel co-integration models. If the observed variables are, for instance, integrated with order of integration one, the convergence rate in Lemma \ref{theo1} will be different (in general faster),  but the shrinkage idea and the consistency of the jumping slope estimator remain valid. All we need is an appropriate threshold that asymptotically dominates the supremum estimator of the zero coefficients, but without being dominated by the estimators for the non-zero coefficients as $n$ and/ or $T$ go to infinity. 
\end{rmk}

\section{Post-SAW Estimation Procedure}\label{postSAWproced} 
\subsection{Tree-Structured Representation}

\begin{figure}[b!]
  \centering
  \includegraphics[width=1\textwidth, trim=10 80 0 80, clip]{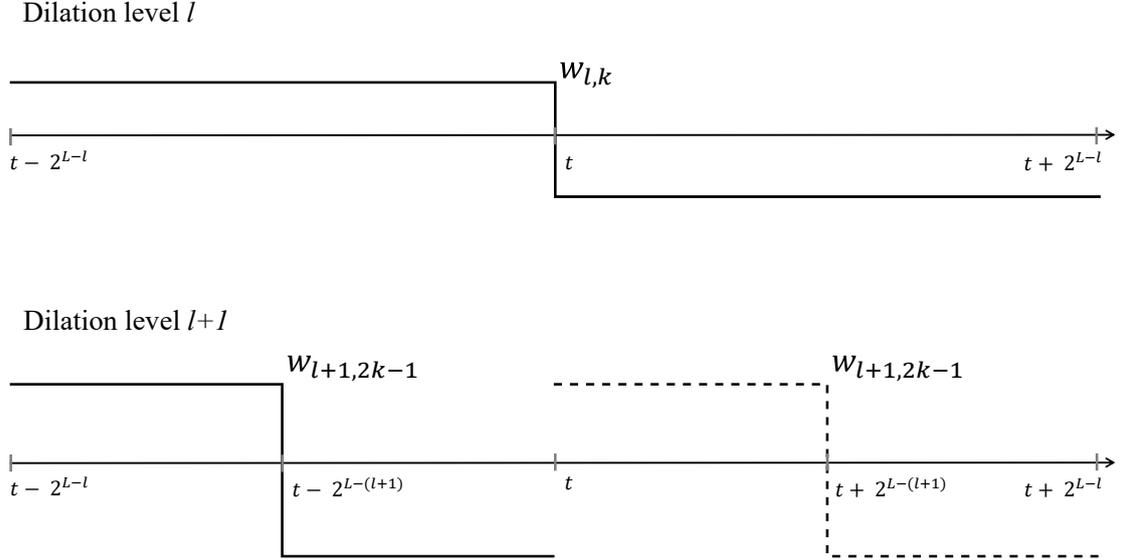}\caption{Classical wavelet basis functions in a parent-child  representation for the levels $l$ and $l+1$.}
  \label{fig:child_parent_wavelet_basis}%
\end{figure}

Because of the special structure of the wavelet basis functions defined in the
interval $t = 1, \ldots, T-1$ by the dyadic dilations $l = 1, \ldots, L$ and the
translations $k = 1, \ldots, K_{l}$, the presence of disturbance errors in the
regression function can induce spuriously estimated jumps, especially when the
true jumps are at non-dyadic positions, which we need to detect and correct. To
understand this phenomenon, we introduce in this section a \textit{tree-structured wavelet representation} which simplifies decoding the role of the detected non-zero wavelet coefficients in approximating the target jumping function.

\begin{figure}[t!]
  \centering
  \includegraphics[width=1\textwidth, trim=10 80 0 80, clip]{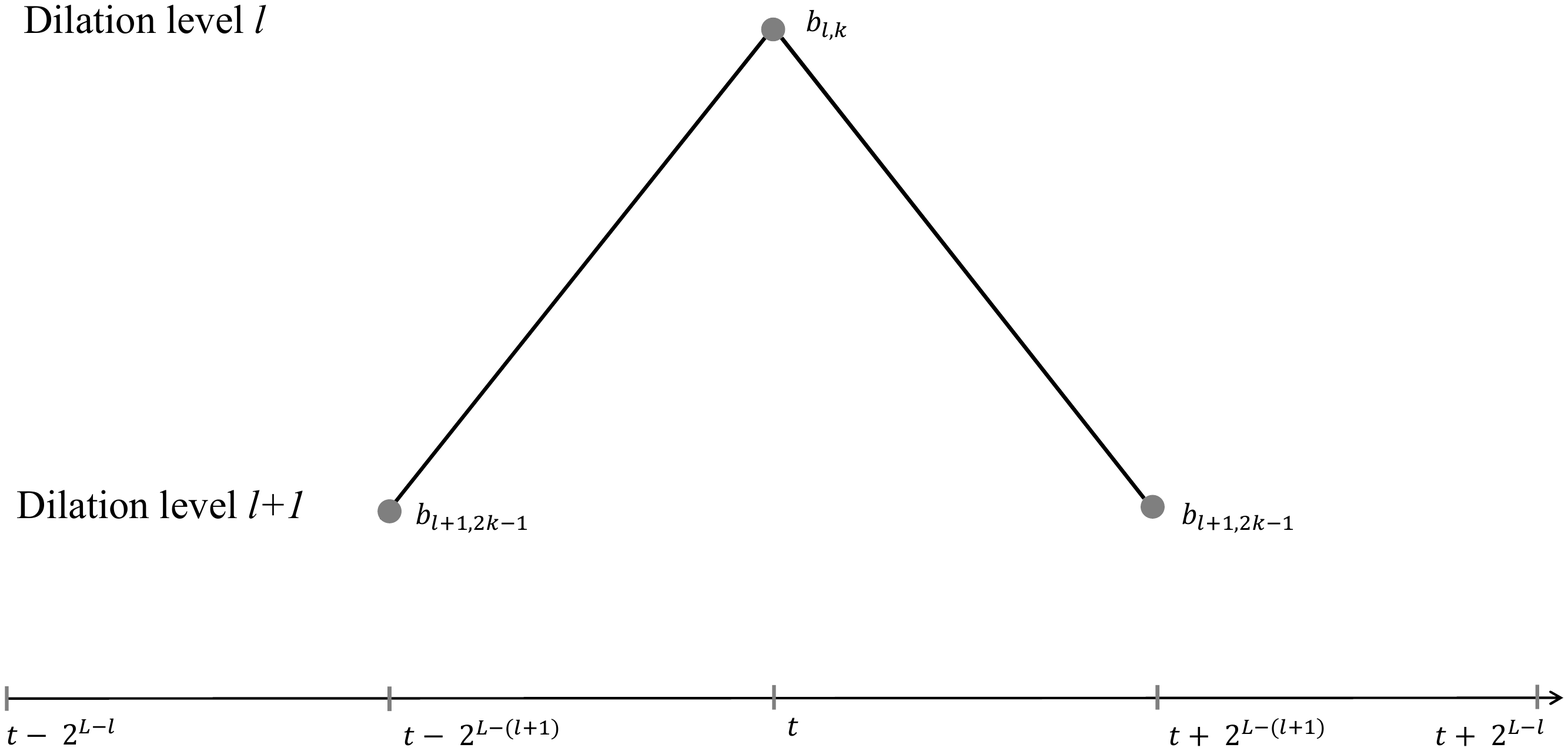}\caption{Wavelet coefficients in a parent - child  representation for the levels $l$ and $l+1$.}
  \label{fig:child_parent_wavelet_coefficient}%
\end{figure}

Recall that, by construction, wavelet basis functions are nested over a binary multiscale structure so that the support interval of the $(l,k)$th basis function comprises the union of the disjunctly halved and translated supports of the basis functions $(l+1,2k-1)$ and $(l+1,2k)$; see Figure \ref{fig:child_parent_wavelet_basis}.

\begin{figure}[ht]
\centering
\includegraphics[width=1\textwidth, trim=10 80 0 80, clip]{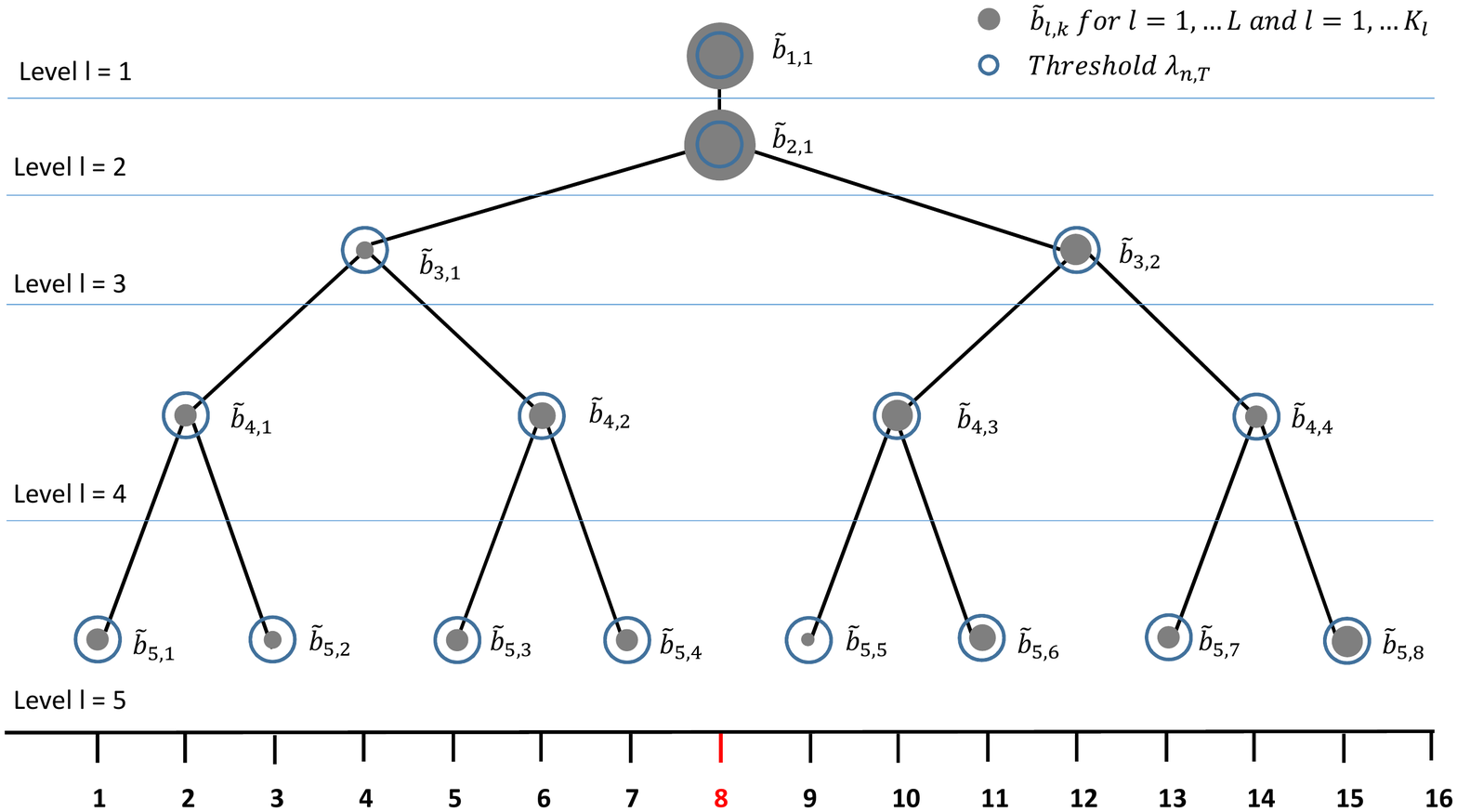}\caption{Example 1: An illustrating example of a tree-structured representation for the wavelet coefficients where the true slope function has a unique jump at a dyadic position $\tau = 8$.}
\label{fig:tree_structured_coef_1}%
\end{figure}

\begin{figure}[ht]
\centering
\includegraphics[width=1\textwidth, trim=10 80 0 80, clip]{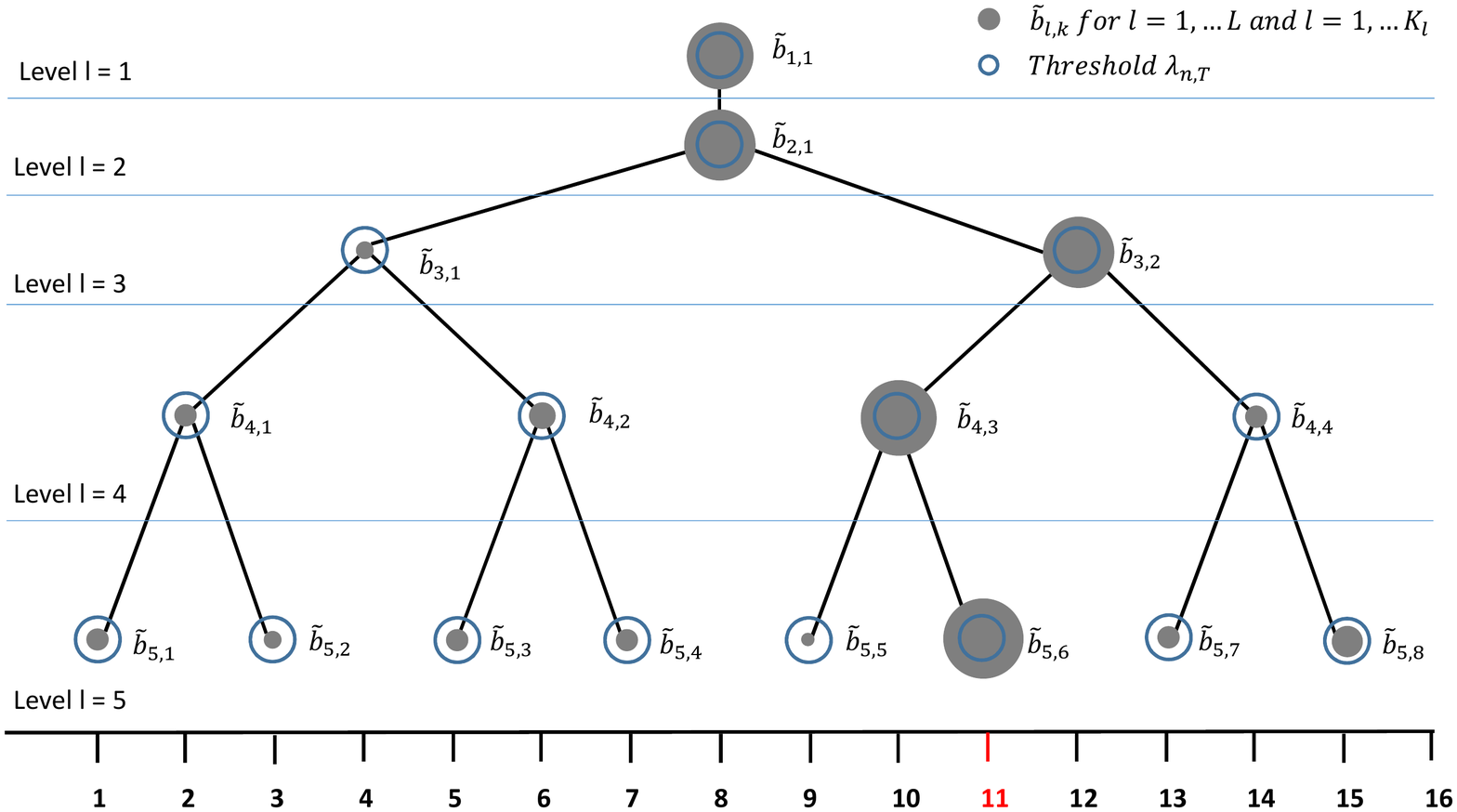}\caption{Example 2: An illustrating example of a tree-structured representation for the wavelet coefficients where the true slope function has a unique jump at a non-dyadic position $\tau = 11$}
\label{fig:tree_structured_coef_2}%
\end{figure}

\begin{figure}[ht]
\centering
\includegraphics[width=1\textwidth, trim=10 80 0 80, clip]{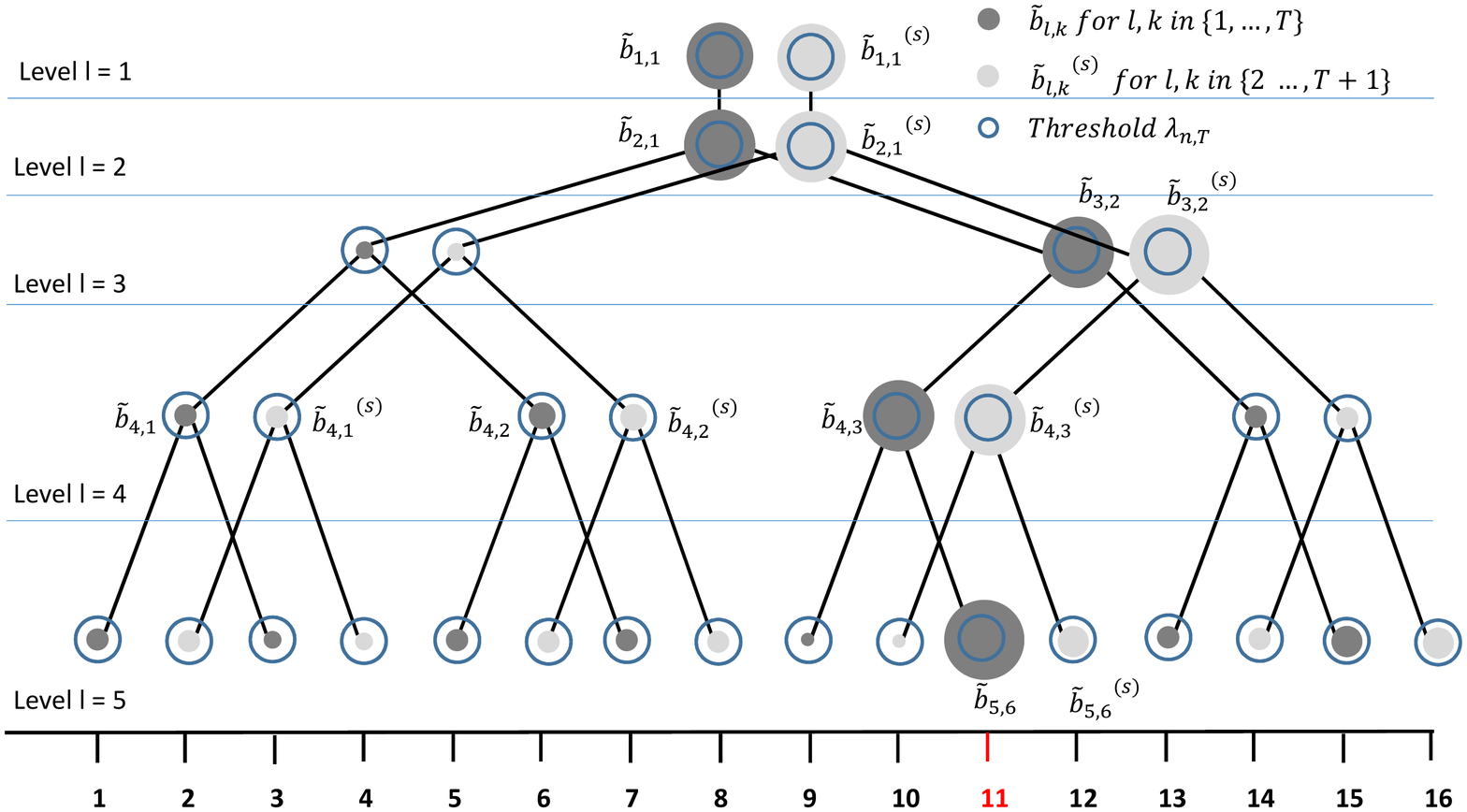}\caption[Tree-structured
representation of the shifted and non-shifted coefficients]{An illustrating
example of a tree-structured representation for the shifted and non-shifted
coefficients.}%
\label{fig:shifted_tree_structured_2}%
\end{figure}

We say that the wavelet coefficient $b_{l,k}$ of the basis function $w_{l,k}$ is
the \textit{parent} of the two \textit{child} coefficients $b_{l+1,2k-1}$ and
$b_{l+1,2k}$ of $w_{l+1, 2k-1}$ and $w_{l+1,2k}$, respectively. This motivates a
hierarchical tree-based representation of the coefficients $b_{l,k},
b_{l+1,2k-1}$ and $b_{l+1,2k}$ in the interval $t - 2^{L-l}, \ldots, t +
2^{L-l}$ as shown in Figure \ref{fig:child_parent_wavelet_coefficient}. Starting
from the highest dilation level $L$ and rooting recursively each wavelet
coefficient to its parent,  we get the whole dyadic tree-structured wavelet
representation as shown in Figures \ref{fig:tree_structured_coef_1}-\ref{fig:shifted_tree_structured_2}.

To encode the potential systematic jumps, we have now to traverse the tree starting from the non-zero coefficients at the finest resolutions (i.e., the highest dilation levels containing estimated coefficients larger than the threshold) up to the primary root parent in a recursive trajectory. All coefficients in these trajectories could be active and potential candidates for spurious jumps at their corresponding coordinates on the time axis, especially when the support of the stability interval of the target function is larger than the support of the basis functions at the detected highest dilation levels. 

As an illustration, consider the two tree-structured representations depicted in
Figure \ref{fig:tree_structured_coef_1} and \ref{fig:tree_structured_coef_2} for
univariate regressions, where $T-1 = 16$. The blue circles in the graphics
represent the height of the threshold $\lambda_{nT}$. The estimated wavelet
coefficients, which are in absolute values $|\tilde{b}_{l,k}|$ smaller than the
threshold $\lambda_{nT}$, are represented by small points randomly sized inside
the circles and represent those coefficients that are shrunk to zero in our
approach. Likewise, the detected non-zero coefficients are represented by oversized points
larger than the threshold circles.

In the first example (see Figure
\ref{fig:tree_structured_coef_1}), we assume that the true univariate slope
function has a unique jump at a perfect dyadic position $\tau = (T-1)/2 = 8$.
Since the highest level of the estimated non-zero coefficients is $l = 2$ and
the corresponding coefficient $\tilde b_{2,1}$ has only one parent, which is the
coefficient of the global constant function, the problem of detecting spurious
jumps does not arise in this simple case.

The second example considers the case of a slope function with one jump at a
non-dyadic location $\tau = 11$ (see Figure \ref{fig:tree_structured_coef_2}).
The highest level containing the non-zero coefficient is $l = 5$ and the
corresponding coefficient $\tilde b_{5,6}$ has three parents, namely $\tilde
b_{4,3}, \tilde b_{3,2},$ and $\tilde b_{2,1}$ excluding the obligatory primary
parent $\tilde b_{1,1}$. Due to the presence of the stochastic error in the
regression function, these estimates deviate randomly from their exact true
values and could generate systematic spurious jumps at  $t = 8,10, 12$.

Theorem \ref{theobetaSAW} shows that the existence of such potential artifacts
does not affect the mean square consistency of our SAW-estimators. However, if
we are interested in the exact locations of the true jumps, we need to go deeper
into the structure of the wavelet coefficients of $\gamma_t$ (remember that
$\gamma_{p,t} = \beta_{p,t}$ and $\gamma_{p+P, t} = \beta_{p,t-1}$ for $p = 1,
\ldots, P$).

Note that the coefficients of the highest level $L$ in the interval $t = 1, \ldots, T-1$ account for the behavior of the slope function at the $(T-1)/2$ odd observations $t = 1,3,5, \ldots, T-2$. If we now redefine the univariate wavelet basis functions in the lagged interval $t = 2, \ldots, T$ instead of $t = 1, \ldots, T-1$, the coordinates of the corresponding wavelet coefficients in the tree-structured representation will be shifted by one time observation, so that the coefficients of the highest level $L$, have to depict the behavior of the slope function at the complement of the odd observation set, that is the even set $\{2, 4, \ldots, T-1\}$.

Continuing with the same example of Figure \ref{fig:tree_structured_coef_2}, we can see that the tree-structured representations of the shifted and un-shifted coefficients does not support the hypothesis of existing additional jumps at t = $8,10, 12$; see Figure \ref{fig:shifted_tree_structured_2}. This is because all coefficients of the shifted tree at the highest level $l = L$ are, as expected, smaller than the threshold.

In the multivariate case, the interpretation of the tree-structured representation can be more complicated since the nodes represent vectors that contain simultaneous information about multiple regressors. In order to construct an individual tree for each parameter, we can re-transform each element of the $\underline P \times1$ vector $\gamma_{t}$ with the conventional univariate wavelet basis functions defined in \eqref{varphi}. Recall that, in our differenced model, ${\gamma}_{t,p} = \beta_{t,p}$ and ${\gamma}_{t,p+P} = \beta_{t-1,p}$. This allows us to obtain for each slope parameter, $\beta_{p}$, two sets of univariate wavelet coefficients:
\begin{align}
\label{newwaveletcoef}c_{l,k,p}^{(s)}  &  = \frac{1}{T-1} \sum_{t = 2}^{T}\psi_{l,k}(t-1) {\gamma}_{t,p},
\end{align}
and
\begin{align}
\label{newwaveletcoefshifted}c_{l,k,p}^{(u)}  &  = \frac{1}{T-1} \sum_{t=1}^{T-1}\psi_{l,k}(t) {\gamma}_{t+1,p+P}.
\end{align}
We use the superscripts $(s)$ and $(u)$ in \eqref{newwaveletcoef} and \eqref{newwaveletcoefshifted} to denote the shifted and un-shifted coefficients, respectively.

Replacing ${\gamma}_{t,p}$ with $\tilde{\gamma}_{t,p} = \sum_{l=1}^{L}\sum_{k=1}^{K_{l}}\sum_{q=1}^{\underline P} W_{lk,p,q}(t)\tilde{{b}}_{l,k,q}$ and ${\gamma}_{t+1,p+P}$ with $\tilde{\gamma}_{t+1,p+P} = \sum_{l=1}^{L}\sum_{k=1}^{K_{l}}\sum_{q=1}^{\underline P} W_{lk,p+P,q}(t+1)\tilde{{b}}_{l,k,q}$, we obtain
\begin{align}
\label{newestiwaveletcoef}\tilde c_{l,k,p}^{(s)}  &  = \frac{1}{T-1} \sum_{t =
2}^{T}\psi_{l,k}(t-1) \tilde{\gamma}_{t,p},
\end{align}
and
\begin{align}
\label{newestiwaveletcoefshifted}\tilde c_{l,k,p}^{(u)}  &  = \frac{1}{T-1}
\sum_{t = 1}^{T-1}\psi_{l,k}(t) \tilde{\gamma}_{t+1,p+P}.
\end{align}
Having an appropriate threshold for $\tilde c_{l,k,p}^{(u)}$, we can construct
the shifted and un-shifted tree-structured representation for each parameter, as
before. This provides then important information about the potential spurious
jumps since all low level parameters in the shifted tree fall in the highest
level of the non-shifted tree and vice versa. Using this insight, we propose a
selection method that allows for a consistent detection of the true jump
locations.  All we need is an appropriate threshold for the highest
coefficients. The following Lemma establishes the uniform consistency in $k$ and $p$ of both
$\tilde{c}^{(s)}_{L,k,p}$ and $\tilde{c}^{(u)}_{L,k,p}$ and states their order
of magnitude in probability. 

\begin{lemma}\label{newlemma2cL}
Suppose Assumptions A, B, and C hold, then, for all $p=1,\ldots,P$ and $m \in \{s,u\}$
$$\sup_{k}   \left|\tilde{c}^{(m)}_{L,k,p} - {c}^{(m)}_{L,k,p} \right| = O_p \left(\log(T-1)^{1/\delta_2}\sqrt{1/n(T-1)}\right).$$
\end{lemma}

\subsection{Detecting the Jump Locations}\label{detlectjumploca}
As described above, interpreting all jumps of the SAW estimator as structural breaks may lead to an over-specification of the break points by including spurious parameter jumps.  In this section, we exploit the information from the shifted and unshifted univariate wavelet coefficients \eqref{newestiwaveletcoef} and \eqref{newestiwaveletcoefshifted} to construct a consistent selection method for detecting the jump locations.

We use \eqref{newestiwaveletcoef} and \eqref{newestiwaveletcoefshifted} to
obtain the following two estimators of $\Delta \beta_{t}$:
\begin{equation}
\Delta \tilde \beta_{t,p}^{(u)} = \sum_{k=1}^{K_{L}}\Delta \psi_{L,k}(t)\hat
c_{L,k,p}^{(u)},\; \text{ for } t \in \mathcal{E},
\end{equation}
and
\begin{equation}
\Delta \tilde \beta_{t,p}^{(s)} = \sum_{k=1}^{K_{L}}\Delta \psi_{L,k}(t-1)\hat
c_{l,k,p}^{(s)},\; \text{ for } t \in \mathcal{E}^{c},
\end{equation}
where
\[
\hat c_{l,k,p}^{(.)} = \tilde c_{l,k,p}^{(.)} \mathbf{I}(|\tilde c_{l,k,p}^{(.)}| >\lambda_{nT}),
\]
$\mathcal{E}$ is the set of the even time locations $\{2, 4, \ldots, T-1\}$,
$\mathcal{E}^{c}$ is the complement set composed of the odd time locations
$\{2,3,4,\ldots, T\} \setminus \mathcal{E}$, and $\mathbf{I}(.)$ is the
indicator function.

The number of jumps of each parameter can be estimated by
\begin{equation}
\label{tildeSppost}\tilde{S}_{p} = \sum_{t \in \mathcal{E}} \mathbf{I}%
(\Delta \tilde \beta_{t,p}^{(u)} \neq0 ) + \sum_{t \in \mathcal{E}^{c}}
\mathbf{I}(\Delta \tilde \beta_{t,p}^{(s)} \neq 0 ).
\end{equation}

The jump locations $\tilde \tau_{1,p}, \ldots, \tilde \tau_{\tilde{S}_{p}, p}$ can be identified as follows:
\begin{equation}
\label{tildtaujppost}\tilde \tau_{j,p} = \min \left \{  s \left|  j = \sum
_{t=2}^{s} \mathbf{I}\left(  \Delta \tilde \beta_{t,p}^{(u)} \neq0 , \, t
\in \mathcal{E} \right)  + \sum_{t=3}^{s} \mathbf{I}\left(  \Delta \tilde
\beta_{t,p}^{(s)} \neq0, \, t \in \mathcal{E}^{c} \right)  \right.  \right \}  ,
\end{equation}
for $j=1,\ldots,\tilde{S}_{p}$.  The maximal number of breaks $S=\sum_{p=1}^{P} S_{p}$ can be estimated by $\tilde S = \sum_{p=1}^{P}\tilde S_{p}$.

\paragraph*{\textbf{Assumption D}: }
\label{assumptionD}
There exists a $c > 0$ such that
 $$\lim_{n, T \to \infty } \big({n}/{\log(T-1)^{2/\delta_2}} \big)^{\kappa^{*} / 2} \min_{j,p}|\beta_{\tau_{j,p}} - \beta_{\tau_{j-1,p}}| > c $$
for any  $0  <\kappa^{*}  < \kappa$. 

Under the asymptotic regime of Assumption A, Assumption D allows the smallest jump size $\min_{j,p}|\beta_{\tau_{j,p}} -
\beta_{\tau_{j-1,p}}|$ to decrease to zero as $n,T \to \infty$, but with a rate slower rate than $\big({n}/{\log(T-1)^{2/\delta_2}} \big)^{-\kappa / 2} $.

\begin{theorem}\label{theo2}
Suppose Assumptions A-D hold, then
\begin{itemize}
\item [(i)] $\lim_{n,T\to \infty} P(\tilde S_1 = S_1, \ldots, \tilde S_p = S_p) = 1$ and
\item [(ii)] $\lim_{n,T \to \infty} P(\tilde \tau_{1,1} = \tau_{1,1},\ldots, \tilde \tau_{S_P, P} = \tau_{S_P, P}|\tilde S_1 = S_1, \ldots, \tilde S_p = S_p) = 1$
\end{itemize}
if $\sqrt{n(T-1)}/\log(T-1)^{1/\delta_2} \lambda_{nT} \to \infty$ and $\sqrt{T-1}\lambda_{nT} \to 0$ as $n,T \to \infty$.
\end{theorem}

Lemma \ref{newlemma2cL} and Theorem \ref{theo2}  justify that asymptotically $\tilde{c}_{L,k,p}^{(m)}$ and $\tilde{b}_{L,k,p}$ can be shrunk by the same threshold $\lambda_{nT}$ introduced in Section \ref{UnivPanMod}. In our simulations we use the following threshold:

\begin{equation}
\label{threshgood2}
\lambda_{nT} = \hat V_{nT}^{1/2} \left(\frac{2\underline{P}\log\big((T-1)\underline{P}\big))}{ n(T-1)^{1/\kappa}}\right)^{{\kappa}/{2}},
\end{equation}

where $\hat V_{nT}$ is the empirical variance estimator corresponding to the largest variance of $(n(T-1))^{-1/2}\sum_{i = 1}^{n} \sum_{t = 2}^{T}\mathcal{Z}_{it,l,k,p}\Delta e_{it}$ over $l,k$, and $p$. Such an estimator can be obtained by using the residuals $\tilde e_{it}$ of a pre-intermediate SAW regression performed with a plug-in threshold $\lambda^{*}_{nT} = 0$.  We want to emphasize that asymptotically we only need that $\hat V_{nT}$ is strictly positive and bounded.  The role of $2 \hat V_{nT}^{1/2}\underline{P}$ is only to give the threshold a convenient amplitude in practice.  As an ad-hoc choice of $\kappa$ we propose using $\kappa=1-\log \log(n(T-1))/\log(n(T-1))$.  Our simulations show that this threshold performs very well.  For more accurate choices of $\kappa$, we refer to the calibration strategies proposed by \citet{Hallin2007} and \citet{Alessi2010}.

\subsection{Post-SAW Estimation}
\label{postsawestimationvk}
For known $\tau_{1,p}, \ldots, \tau_{S_{p},p}$, we can rewrite Model \eqref{diffeq} as
\begin{equation}
\label{postwavaa}
\Delta Y_{it} = \sum_{p = 1}^{P} \sum_{j = 1}^{S_{p} +
1}  X_{it,p}^{(\tau_{j,p})} \beta_{\tau_{j,p}}  - \sum_{p = 1}^{P} \sum_{j = 1}^{S_{p} +
1}  X_{i,t-1,p}^{(\tau_{j,p})} \beta_{\tau_{j,p}}  + \Delta  \theta_t + \Delta  e_{it},
\end{equation}
where
\[
 X_{it,p}^{(\tau_{j,p})} = X_{it,p}\mathbf{I}%
\big(\tau_{j-1,p} < t \leq \tau_{j,p}\big),
\]
with $\tau_{0,p} = 0$ and $\tau_{S_{p} +1,p} = T$, for $p=1,\ldots,P$.

In order to eliminate the effects of $ \Delta  \theta_t$ one usually relies on averaging over individuals:
\begin{align}
\Delta \dot Y_{it}& = \sum_{p = 1}^{P} \sum_{j = 1}^{S_{p} +
1} \dot X_{it,p}^{(\tau_{j,p})} \beta_{\tau_{j,p}} -  \sum_{p = 1}^{P} \sum_{j = 1}^{S_{p} +
1} \dot X_{i,t-1,p}^{(\tau_{j,p})} \beta_{\tau_{j,p}}+ \Delta \dot e_{it}.\notag\\
& = \sum_{p = 1}^{P} \sum_{j = 1}^{S_{p} + 1}\Delta \dot X_{it,p}^{(\tau_{j,p})} \beta_{\tau_{j,p}} + \Delta \dot e_{it},\label{postwav}
\end{align}
Here, the dot operator transforms the variables as follows: $\dot u_{it} = u_{it} - \frac{1}{n} \sum_{i=1}^{n} u_{it}$.

Let $\beta_{(\tau)} = (\beta_{\tau_{1,1}}, \allowbreak \ldots, \allowbreak
\beta_{\tau_{S_{1}+1,1}}, \allowbreak \ldots, \allowbreak \beta_{\tau_{1,P}}
\allowbreak \ldots, \allowbreak \beta_{\tau_{S_{P}+1,P}})^{{\prime}}$ the $D$-dimensional vector of the slope parameters, where $D = P + \sum_{p=1}^{P}S_{p}$. Defining a corresponding $D$-dimensional vector
$$\tilde{\Delta} \dot X_{it,(\tau)}^{'}=
(\Delta \dot X_{it,1}^{(\tau_{1,1})}, \dots, \Delta \dot X_{it,1}^{(\tau_{S_1+1,1})},\dots,
 \Delta \dot X_{it,P}^{(\tau_{1,P})}, \dots, \Delta \dot X_{it,P}^{(\tau_{S_P+1,P})}),$$
we obtain in vector notation
\begin{equation}
\Delta \dot{Y}_{it} = \tilde{\Delta} \dot{X}_{it,(\tau)}^{^{\prime}} \beta_{(\tau)} +
\Delta \dot{e}_{it},
\end{equation}

Let now $Z_{it,(\tau)}$ be a $D$-dimensional vector of instruments appropriate for $\tilde{\Delta} \dot X_{it,(\tau)}$.
For known jump locations, the (infeasible) conventional IV estimator of
$\beta_{(\tau)}$ is
\begin{equation}
\label{convIV}\hat \beta_{(\tau)} = \big(\sum_{i=1}^{n} \sum_{t = 2}^{T}
Z_{it,(\tau)} \tilde\Delta \dot{X}_{it,(\tau)}^{^{\prime}}\big)^{-1} \big(\sum
_{i=1}^{n} \sum_{t = 2}^{T} Z_{it,(\tau)} \Delta \dot{Y}_{it} \big).
\end{equation}

Our final estimator relies on replacing the true jump locations in \eqref{convIV} with the detected jump locations $\tilde \tau:= \{ \tilde \tau_{j,p}\; j=1,\ldots,S_{p}+1,\; p=1,\ldots,P\}$. Our post-SAW estimator is hence

\begin{equation}
\label{postwevIV}\hat \beta_{(\tilde \tau)} = \big(\sum_{i=1}^{n} \sum_{t =
2}^{T} Z_{it,(\tilde \tau)} \tilde\Delta \dot{X}_{it,(\tilde \tau)}^{^{\prime}%
}\big)^{-1} \big(\sum_{i=1}^{n} \sum_{t = 2}^{T} Z_{it,(\tilde \tau)}
\Delta \dot{Y}_{it} \big).
\end{equation}

From \eqref{multempIV} and \eqref{postwevIV}, we can see that the number of
parameters to be estimated after detecting the jump locations is much smaller
than the number of parameters required to estimate the slope parameters in the
SAW regression ($\sum_{p = 1}^{P} (S_{p} + 1) < T(\underline P- 1)$). It is
evident that such a gain in terms of regression dimension improves the quality
of the estimator.

\paragraph*{\textbf{Assumption E} - Central limits:}

Let $\mathcal{T}_{(\tau)}$ be a $D\times D$ diagonal matrix with the diagonal elements ${\ T_{1,1}},
\allowbreak \ldots, \allowbreak{\ T_{S_{P}+1,P}}$, where $T_{j,p} = \tau_{j,p}
- \tau_{j-1,p} + 1$ for $ j=1,\ldots,S_p$ and $T_{S_p+1,p} = \tau_{S_p+1,p}
- \tau_{S_p,p}$ and $\tau_{j,p} - \tau_{j-1,p} \to \infty$ for $j = 1, \ldots, S_p +1$.

\begin{itemize}
\item[$(i):$] $(n \mathcal{T}_{(\tau)})^{-1} \sum_{i = 1}^{n} \sum_{t = 2}^{T}
Z_{it,(\tau)}\tilde\Delta \dot{X}_{it, (\tau)}^{^{\prime}} \overset{p}{\rightarrow}
Q^{\circ}_{(\tau)}$ where $Q^{\circ}_{(\tau)}$ is a full rank finite matrix.

\item[$(ii):$] $(n \mathcal{T}_{(\tau)})^{-1} \sum_{i = 1}^{n} \sum_{t =
2}^{T} \sum_{j = 1}^{n} \sum_{s = 2}^{T} Z_{it,(\tau)}Z_{js,(\tau)}^{^{\prime
}} \sigma_{ij,ts} \overset{p}{\rightarrow} V^{\circ}_{(\tau)}$, where
$V^{\circ}_{(\tau)}$ is a full rank finite matrix.

\item[$(iii):$] $\big(n\mathcal{T}_{(\tau)}\big)^{-\frac{1}{2}} \sum_{i=1}^{n}
\sum_{t = 2}^{T} Z_{it,(\tau)} \Delta \dot{e}_{it} \allowbreak \overset
{d}{\rightarrow} \allowbreak N(0, V^{\circ}_{(\tau)})$.
\end{itemize}

Assumption E presents standard assumptions that are commonly used in the
literature on instrumental variables.

\begin{theorem}[Post-SAW estimator]\label{theo4}
Suppose Assumptions A-E hold. Then conditional on $\tilde S_1 = S_1, \ldots, \tilde S_P = S_P$, we have
\[
\sqrt{n\mathcal{T}_{(\tau)}}\big(\hat \beta_{(\tilde \tau)} -  \beta_{(\tau)} \big) \stackrel{d}{\rightarrow} N(0, \Sigma_{(\tau)}),
\]
where
$\Sigma_{(\tau)} = ({Q}^{\circ}_{(\tau)})^{-1} \big( V^{\circ}_{(\tau)} \big) ({Q}^{\circ}_{(\tau)})^{-1}$.
\end{theorem}If $T\rightarrow \infty$ and all $T_{j,p}$ diverge proportionally
to $T$, then $\hat{\beta}_{\tau_{j},p}$ achieves the usual $\sqrt{nT}$-
convergence rate.
According to Theorem \ref{theo4}, our final estimator of the model parameters are
first-order efficient: they have the same asymptotic distribution as the
(infeasible) estimators that would be obtained if all jump locations were
exactly known a priori and thus possess the "oracle property" in regard to
these parameter estimates. The final estimator proposed by \citet{Qian2014}
also shares this oracle property.

To summarize, our method essentially consists of two steps. A structure adapted wavelet (SAW) estimation followed by the post-SAW procedures. Our SAW method could be applied in a time series context by fixing $n$ and letting $T$ be large, in which case our approach may be considered a more parsimonious variant of the subinterval approach of \citet{Bai1998, Bai2003a}. Although we do not pursue this issue in this paper we do show in Theorem 1 that our wavelet procedure provides (mean square) consistent parameter estimates even for fixed $n$ and large $T$. In contrast, the method of \citet{Qian2014} essentially consists of three steps. In a first step they use the ``na\"{\i}ve'' estimator of $\beta_{t}$ for every single time point $t$. This is used in the second step to construct appropriate weights for the adaptive lasso procedure. In a third step they then propose a post lasso instead of our wavelet procedure. An important point is that their method will only work if $n$ is large. Their method does not consider cases in which $n$ is small and $T$ is large as we do.

Because the asymptotic variance $\Sigma_{(\tau)}$ of $\hat \beta_{(\tilde \tau)}$ in Theorem  \ref{theo4} is unknown, consistent estimators of $Q^{\circ
}_{(\tau)}$ and $V^{\circ}_{(\tau)}$ are required to perform inferences. A
natural estimator of $Q^{\circ}_{(\tau)}$ is
\[
\hat Q_{(\tilde \tau)} = (n \mathcal{T}_{(\tilde \tau)})^{-1} \sum_{i = 1}^{n}
\sum_{t = 2}^{T} Z_{it,(\tilde \tau)}\tilde \Delta \dot{X}_{it, (\tilde \tau)}^{^{\prime}}%
\]
and a consistent estimator of $\Sigma_{(\tau)}$ can be obtained by
\[
\hat \Sigma_{(\tilde \tau), j} = \hat Q_{(\tilde \tau)}^{-1}\hat V_{(\tilde \tau
)}^{(c)} \hat Q_{(\tilde \tau)}^{-1},
\]
where $\hat V_{(\tilde \tau)}^{(c)}$ a consistent estimator of $V_{(\tilde
\tau)}^{\circ}$ that can be constructed depending on the structure of
$\Delta \dot{e}_{it}$. For brevity, we distinguish only four cases:

\begin{enumerate}
\item The case of homoscedasticity without the presence of auto- and
cross-section correlations:
\[
\hat V_{(\tilde \tau)}^{(1)} = (n \mathcal{T}_{(\tilde \tau)})^{-1} \sum_{i =
1}^{n} \sum_{t = 2}^{T} Z_{it,(\tilde \tau)}Z_{it,(\tilde \tau)}^{^{\prime}}%
\hat \sigma^{2},
\]
where $\hat \sigma^{2} = \frac{1}{n(T-1)}\sum_{i=1}^{n}\sum_{t=2}^{T}
\Delta \hat{\dot{e}}_{it}^{2}$, with $\Delta \hat{\dot{e}}_{it} = \Delta \dot
{Y}_{it} - \Delta \dot{X}_{it,(\tilde \tau)}^{^{\prime}} \hat \beta_{(\tilde
\tau)}$.

\item The case of cross-section heteroscedasticity without auto- and
cross-section correlations:
\[
\hat V_{(\tilde \tau)}^{(2)} = (n \mathcal{T}_{(\tilde \tau)})^{-1} \sum_{i =
1}^{n} \sum_{t = 2}^{T} Z_{it,(\tilde \tau)}Z_{it,(\tilde \tau)}^{^{\prime}}%
\hat \sigma^{2}_{i},
\]
where $\hat \sigma^{2}_{i} = \frac{1}{T-1}\sum_{t=2}^{T} \Delta \hat{\dot{e}%
}_{it}^{2}$.

\item The case of time heteroscedasticity without auto- and cross-section
correlations:
\[
\hat V_{(\tilde \tau)}^{(3)} = (n \mathcal{T}_{(\tilde \tau)})^{-1} \sum_{i =
1}^{n} \sum_{t = 2}^{T} Z_{it,(\tilde \tau)}Z_{it,(\tilde \tau)}^{^{\prime}}%
\hat \sigma^{2}_{t},
\]
where $\hat \sigma^{2}_{t} = \frac{1}{n}\sum_{i=1}^{n} \Delta \hat{\dot{e}}%
_{it}^{2}$.

\item The case of cross-section and time heteroscedasticity without auto- and
cross-section correlations:
\[
\hat V_{(\tilde \tau)}^{(4)} = (n \mathcal{T}_{(\tilde \tau)})^{-1} \sum_{i =
1}^{n} \sum_{t = 2}^{T} Z_{it,(\tilde \tau)}Z_{it,(\tilde \tau)}^{^{\prime}%
}\Delta \hat{\dot{e}}_{it}^{2}.
\]

\end{enumerate}

\begin{proposition} \label{varianceestim}
Under Assumptions A-E, we have, as $n,T \to \infty$, $
\hat \Sigma_{(\tilde \tau)}^{(c)} = \Sigma_{(\tau)} + o_p(1)$, for $c = 1,2,3,$ and $4$. 
\end{proposition}

Based on the asymptotic distribution of $\hat{\beta}_{(\tilde{\tau})}$ in Theorem \ref{theo4} the variance estimators presented in Proposition \ref{varianceestim}, usual asymptotic test statistics such as $z$-tests and $\chi^2$-tests can be used for inference.

\begin{rmk}
If the errors (at the difference level) are autocorrelated,  $V_{(\tilde \tau)}^{(c)}$ can be estimated by applying the standard heteroscedasticity and autocorrelation (HAC) robust limiting covariance estimator to the sequence $\{Z_{it, (\tilde \tau)}\Delta \hat{\dot{e}}_{it}\}_{i, t}$ for $i\in \mathbb{N}^{*}$ and $t\in \mathbb{N}^{*}\setminus \{1\}$; see, e.g., \citet{Newey87}. In the presence of additional cross-section correlations, one can use the partial sample method together with the Newey-West procedure as proposed by \citet{Bai2009a}. A formal proof of consistency remains, in this case, to be explored.
\end{rmk}

\begin{rmk}
In some particular applications researchers may need to impose the restriction of contemporaneous parameter jumps for all regressors.  Such a restriction can be easily implemented by using the union of all detected jump-locations, $\bigcup_{p=1}^P\bigcup_{j=1}^{S_p}\tau_{j,p}$, for each regressor in the post-SAW estimator (29).  This adapted estimation procedure is then consistent, given the assumption of common jump-locations.
\end{rmk}

\section{Monte Carlo Simulations}\label{sumulations}
\begin{table}[b!]
\setlength{\tabcolsep}{4pt}
\small\centering
\begin{tabular}{rr llll lll}
\toprule
\multicolumn{9}{c}{DGP1 ($S_1=2$ and $S_2=3$)}\\
\midrule
\multicolumn{2}{l}{} & \multicolumn{4}{c}{Post-SAW} & \multicolumn{3}{c}{\cite{Qian2014}} \\
\cmidrule(l{3pt}r{3pt}){3-6} \cmidrule(l{3pt}r{3pt}){7-9}
$T$ &$n$ &$\tilde{S}_1$&$\tilde{S}_2$&HD$_1/T$  &HD$_2/T$    &$\hat{S}_{QS}$& HD$_1/T$ & HD$_2/T$\\

\midrule
   33 &  30 & 2.00 (0.00) & 3.00 (0.00) & 0.00 (0.00) & 0.00 (0.00) & 9.23 (2.99) & 0.23 (0.06) & 0.16 (0.05) \\ 
   33 &  60 & 2.00 (0.00) & 3.00 (0.00) & 0.00 (0.00) & 0.00 (0.00) & 6.07 (1.44) & 0.17 (0.05) & 0.11 (0.04) \\ 
   33 & 120 & 2.00 (0.00) & 3.00 (0.00) & 0.00 (0.00) & 0.00 (0.00) & 5.20 (0.50) & 0.16 (0.02) & 0.10 (0.02) \\ 
   33 & 300 & 2.00 (0.00) & 3.00 (0.00) & 0.00 (0.00) & 0.00 (0.00) & 5.00 (0.00) & 0.15 (0.00) & 0.09 (0.00) \\ 
\midrule
   65 &  30 & 2.00 (0.00) & 3.00 (0.00) & 0.00 (0.00) & 0.00 (0.00) & 6.68 (2.34) & 0.19 (0.06) & 0.12 (0.05) \\ 
   65 &  60 & 2.00 (0.00) & 3.00 (0.00) & 0.00 (0.00) & 0.00 (0.00) & 5.20 (0.56) & 0.16 (0.02) & 0.10 (0.02) \\ 
   65 & 120 & 2.00 (0.00) & 3.00 (0.00) & 0.00 (0.00) & 0.00 (0.00) & 5.02 (0.17) & 0.16 (0.01) & 0.09 (0.01) \\ 
   65 & 300 & 2.00 (0.00) & 3.00 (0.00) & 0.00 (0.00) & 0.00 (0.00) & 5.00 (0.00) & 0.15 (0.00) & 0.09 (0.00) \\ 
\midrule
  129 &  30 & 2.00 (0.00) & 3.00 (0.00) & 0.00 (0.00) & 0.00 (0.00) & 5.24 (0.73) & 0.17 (0.02) & 0.09 (0.02) \\ 
  129 &  60 & 2.00 (0.00) & 3.00 (0.00) & 0.00 (0.00) & 0.00 (0.00) & 5.02 (0.15) & 0.16 (0.01) & 0.09 (0.00) \\ 
  129 & 120 & 2.00 (0.00) & 3.00 (0.00) & 0.00 (0.00) & 0.00 (0.00) & 5.00 (0.00) & 0.16 (0.00) & 0.09 (0.00) \\ 
  129 & 300 & 2.00 (0.00) & 3.00 (0.00) & 0.00 (0.00) & 0.00 (0.00) & 5.00 (0.00) & 0.16 (0.00) & 0.09 (0.00) \\
\bottomrule
\end{tabular}
\caption{Simulation results for DGP1. The table shows the means (sd) computed from the
  500 Monte Carlo replications. Results are rounded to two decimal places.}\label{DGP1}
\end{table}

In this section, we use Monte Carlo simulations to examine the finite sample performance
of our method. Our data generating processes (DGPs) are based on the following general
panel data model

\[
    Y_{it}=\mu + \alpha_i + \theta_t +\sum_{p=1}^PX_{it,p}\beta_{t,p} + \sigma_{it} e_{it}\quad\text{for}\quad
i=1,\dots,n\;\;\text{and}\;\;t=1,\dots T,
\]
where

\begin{equation}\label{eq:beta_MCsim}
\beta_{t,p}=\left \{
\begin{array}
[c]{lcl}%
\beta_{\tau_{1,p}} & \text{for} & t=1,\dots,\tau_{1,p}\\
\;\;\vdots &  & \\
\beta_{\tau_{S_p+1,p}} & \text{for} & t=\tau_{S_p,p}+1,\dots,T
\end{array}
\right.
\end{equation}
with
\[
\beta_{\tau_j, p} = \frac{a_n}{3}\cdot(-1)^{j}\;\;\text{with}\;\;\tau_{j}=\left\lfloor
\frac{j}{S+1}(T-1)\right \rfloor,\;\text{for }j=1,\ldots,S.
\]
With $\beta_{\tau_j, p}$ dependent on $n$ we allow for the signal to decrease as $n$
increases. While the case of $S=1$ represents a dyadic jump location, the cases $S=2$
and $S=3$ represent non-dyadic jump locations.

To see how the properties of the estimators vary with $n$ and $T$, we consider $12$
different sample size combinations spanned by $T=33,65,129$ and $n=30,60,120,300$. We examine in
total six different DGPs. For all DGPs we set $(a_{30}, a_{60}, a_{120}, a_{300}) = (7,
5, 4, 3)$. DGP1 consists of $P=2$ regressors with $S_1=2$ and $S_2=3$ jump locations.
DGP2 considers the case of endogeneous regressors. DGP3 and DGP4 examine the effects of
heteroscedasticity across time and cross-section as well as autocorrelated error terms.
DGP5 includes time fixed effects. DGP2-DGP5 all have one regressor and are analyzed with
$S=1,2,3$ jump locations. Finally, DGP6 investigates the no-jump case ($S=0$). For each
DGP we compare our Post-SAW estimation method with the method of \cite{Qian2014}.  We
report the means and standard deviations computed from 500 Monte Carlo replications of
the following statistics: the number(s) of jump locations, $\tilde{S}$, the scaled and
squared Euclidean distance,
$||\hat\beta_p-\beta_p||^2/T=T^{-1}\sum_{t=1}^T(\hat{\beta}_{t,p}-\beta_{t,p})^2$,
between $\hat\beta_p=(\hat\beta_{1,p},\dots,\hat\beta_{T,p})'$ and
$\beta_p=(\beta_{1,p},\dots,\beta_{T,p})'$, the scaled Hausdorff distance, HD$/T$,
between the estimated, $\hat{\tau}_{j,p}$, and the true jump locations, $\tau_{j,p}$,
where the scaling by $T$ allows us to compare distances for different values of $T$. In
the following, we define the considered DGPs.

\begin{table}[t!]
\small\centering
\begin{tabular}{rr ll ll}
\toprule
\multicolumn{6}{c}{DGP1 ($S_1=2$ and $S_2=3$)}\\
\midrule
\multicolumn{2}{l}{}&\multicolumn{2}{c}{Post-SAW}&\multicolumn{2}{c}{\cite{Qian2014}} \\
\cmidrule(l{3pt}r{3pt}){3-4} \cmidrule(l{3pt}r{3pt}){5-6}
$T$ & $n$ & $||\hat{\beta}_1-\beta_1||^2/T$& $||\hat{\beta}_2-\beta_2||^2/T$ & $||\hat{\beta}_{1}-\beta_{1}||^2/T$ &$||\hat{\beta}_{2}-\beta_{2}||^2/T$\\

\midrule
   33 &  30 & 0.005 (0.004) & 0.007 (0.005) & 0.050 (0.027) & 0.052 (0.027) \\ 
   33 &  60 & 0.002 (0.002) & 0.003 (0.002) & 0.015 (0.010) & 0.014 (0.009) \\ 
   33 & 120 & 0.001 (0.001) & 0.002 (0.001) & 0.005 (0.003) & 0.005 (0.003) \\ 
   33 & 300 & 0.000 (0.000) & 0.001 (0.000) & 0.002 (0.001) & 0.002 (0.001) \\ 
\midrule
   65 &  30 & 0.002 (0.002) & 0.003 (0.002) & 0.019 (0.014) & 0.018 (0.015) \\ 
   65 &  60 & 0.001 (0.001) & 0.002 (0.001) & 0.005 (0.004) & 0.005 (0.004) \\ 
   65 & 120 & 0.001 (0.000) & 0.001 (0.001) & 0.002 (0.001) & 0.002 (0.001) \\ 
   65 & 300 & 0.000 (0.000) & 0.000 (0.000) & 0.001 (0.001) & 0.001 (0.001) \\ 
\midrule
  129 &  30 & 0.001 (0.001) & 0.002 (0.001) & 0.006 (0.004) & 0.006 (0.004) \\ 
  129 &  60 & 0.001 (0.001) & 0.001 (0.001) & 0.002 (0.001) & 0.002 (0.002) \\ 
  129 & 120 & 0.000 (0.000) & 0.000 (0.000) & 0.001 (0.001) & 0.001 (0.001) \\ 
  129 & 300 & 0.000 (0.000) & 0.000 (0.000) & 0.000 (0.000) & 0.000 (0.000) \\ 
\bottomrule

\end{tabular}
\caption{Simulation results for DGP1. The table shows the means (sd) computed from the
500 Monte Carlo replications. Results are rounded to three decimal places.}\label{DGP1b}%
\end{table}

\smallskip

\noindent\textbf{DGP1} (\textit{multiple regressors}): This DGP has $P=2$ regressors,
$X_{it,1}$ and $X_{it,2}$, that are uncorrelated with the error term $e_{it}\sim
N(0,2)$,
\begin{align*}
X_{it,1}= 0.5 \alpha_i + \xi_{it,1}\quad\text{and}\quad X_{it,2}= 0.5 \alpha_i + \xi_{it,2}
\end{align*}
where $\alpha_i$, $\xi_{it,1}$, $\xi_{it,2}\sim N(0,1)$, $\mu=0, \theta_t = 0$, and
$\sigma_{it} = 1$ for all $i$ and $t$.  The slope parameters are defined as in
\eqref{eq:beta_MCsim} where $\beta_{t,1}$ and $\beta_{t,2}$ are using $S_1=2$ and
$S_2=3$ jump locations, respectively.

\bigskip

\noindent\textbf{DGP2} (\textit{endogeneous regressor}): This DGP has one endogeneous
regressor $X_{it}= 3 \cdot Z_{it} + e_{it}$ with instrument $Z_{it} = 0.5 \alpha_{i} +
\xi_{it}$, where $\alpha_i, \xi_{it} \sim N(0,1)$ and $e_{it} \sim N(0, 0.5)$. Again
$\mu=0, \theta_t = 0$ and $\sigma_{it} = 1$ for all $i$ and $t$. The slope parameter is
defined as in \eqref{eq:beta_MCsim} for which we consider three different jump locations
$S=1,2,3$.

\bigskip

\noindent\textbf{DGP3} (\textit{heteroscedasticity in time- and cross-section}): This
DGP has one regressor $X_{it}=0.5 \alpha_{i} + \xi_{it}$ with $\alpha_{i},\xi_{it}\sim
N(0,1)$. The errors are heteroscedastic in both time- and cross-section, with $e_{it}
\sim N(0, 0.5)$ and $\sigma^2_{it} \sim U(1, 3)$. Here $\mu = 0$ and $\theta_t = 0$ for
all $t$. The slope parameter is as in DGP2.

\bigskip

\noindent\textbf{DGP4} (\textit{heteroscedasticity in the coss-section and serial
correlation}): This DGP has the same regressor structure as DGP3. The error term is
autocorrelated with individual specific persistence: $e_{it} = \rho_{i} e_{i,t-1} +
\zeta_{it}$ with $\rho_{i} \sim U(.25,.75)$ and $\zeta_{it}\sim N(0,3)$. As in DGP2
$\mu = 0, \theta_t = 0$ and  $\sigma_{it} = 1$ for all $i$ and $t$. The slope parameter
is as in DGP2.

\bigskip

\noindent\textbf{DGP5} (\textit{heteroscedasticity and time fixed effects}):
This DGP is as DGP3 with $\sigma_{it}^2 \sim U(1, 2)$ but in addition includes a time
fixed effect with jumps.  We define $\theta_t$ to be as $\beta_t$ in
\eqref{eq:beta_MCsim} with $a_n = 7$ for all $n$ and $S_{\theta} = \lfloor T / 10
\rfloor$ many jump locations dependent on the panel length.

\bigskip

\noindent\textbf{DGP6} (\textit{no-jumps}): This DGP is equivalent to DGP4 except that
the slope parameter does not contain structural breaks and the error variance is
increased. We set $\beta_{t} = 1$ for all $t$ and let $\zeta_{it} \sim N(0, 4)$.

\smallskip

\begin{table}[b!]
\small\centering
\begin{tabular}{rr lll lll}
\toprule
\multicolumn{8}{c}{DGP4} \\
\midrule
&&\multicolumn{3}{c}{Post-SAW}&\multicolumn{3}{c}{\cite{Qian2014}}\\
\cmidrule(l{3pt}r{3pt}){3-5} \cmidrule(l{3pt}r{3pt}){6-8}
  $T$&$n$&$\tilde{S}$&$||\hat{\beta}-\beta||^2/T$&HD$/T$&$\hat{S}$&$||\hat{\beta}-\beta||^2/T$&HD$/T$\\
\midrule
  33 & 30  & 2.99 (0.09) & 0.03 (0.24) & 0.00 (0.02) & 7.48 (2.85) & 0.05 (0.03) & 0.15 (0.07) \\ 
  33 & 60  & 2.99 (0.10) & 0.02 (0.14) & 0.00 (0.02) & 4.59 (1.72) & 0.01 (0.01) & 0.08 (0.08) \\ 
  33 & 120 & 3.00 (0.04) & 0.00 (0.04) & 0.00 (0.01) & 3.44 (0.86) & 0.00 (0.00) & 0.03 (0.06) \\ 
  33 & 300 & 3.00 (0.00) & 0.00 (0.00) & 0.00 (0.00) & 3.07 (0.30) & 0.00 (0.00) & 0.01 (0.03) \\ 
\midrule
  65 & 30  & 3.00 (0.04) & 0.01 (0.12) & 0.00 (0.01) & 6.06 (3.31) & 0.02 (0.02) & 0.11 (0.09) \\ 
  65 & 60  & 2.97 (0.18) & 0.05 (0.25) & 0.01 (0.04) & 3.61 (1.16) & 0.00 (0.00) & 0.04 (0.06) \\ 
  65 & 120 & 3.00 (0.04) & 0.00 (0.04) & 0.00 (0.01) & 3.10 (0.36) & 0.00 (0.00) & 0.01 (0.03) \\ 
  65 & 300 & 3.00 (0.00) & 0.00 (0.00) & 0.00 (0.00) & 3.01 (0.08) & 0.00 (0.00) & 0.00 (0.01) \\ 
\midrule
  129 & 30  & 2.98 (0.13) & 0.05 (0.36) & 0.00 (0.03) & 3.77 (1.59) & 0.01 (0.01) & 0.04 (0.07) \\ 
  129 & 60  & 2.97 (0.18) & 0.05 (0.24) & 0.01 (0.04) & 3.14 (0.45) & 0.00 (0.00) & 0.01 (0.04) \\ 
  129 & 120 & 2.99 (0.12) & 0.01 (0.10) & 0.00 (0.03) & 3.02 (0.13) & 0.00 (0.00) & 0.00 (0.01) \\ 
  129 & 300 & 3.00 (0.00) & 0.00 (0.00) & 0.00 (0.00) & 3.00 (0.00) & 0.00 (0.00) & 0.00 (0.00) \\ 
\bottomrule

\end{tabular}
\caption{Simulation results for DGP4 ($S=3$). The table shows the means (sd) computed
from the 500 Monte Carlo replications. Results are rounded to two decimal places.}%
\label{DGP4_S3}
\end{table}

While our SAW estimation procedure allows for different jump locations in each of the
regressors, the method of \cite{Qian2014} assumes a common jump structure for all
regressors. Table \ref{DGP1} demonstrates the effect of this fundamental methodological
difference between the two approaches. The Post-SAW estimator is able to detect the
correct numbers of jumps, $S_1=2$ and $S_2=3$ and the correct jump locations for both
regressors. The method of \cite{Qian2014}, however, is not able to
produce consistent estimation results for the jump locations when the jump locations
differ between the regressors. This inconsistency in estimating the correct break
points leads to more inefficient parameter estimates due to the superfluously estimated
break points (see Table \ref{DGP1b}). The small sample case ($n=30$) demonstrates the
pratical relevance for our theoretical finite $n$ results in Theorem \ref{theobetaSAW}
part \textit{(ii)}.

The estimation results for DGP2-DGP6 generally show the same basic structure. By
    contrast to DGP1, each of these DGPs considers only one regressor such that both
    methods, Post-SAW and the method of \cite{Qian2014}, show comparable estimation
    results for large sample sizes. However, for small samples sizes our Post-SAW
    estimation procedure shows advantages over the method of \cite{Qian2014}.\\
    In Table \ref{DGP4_S3} and Table \ref{DGP5_S3} we show the simulation results for
    DGP4 and DGP5 with $S=3$ break points. DGP4 and DGP5 demonstrate that our
    thresholding approach is robust to cross-section heterogeneity, serial correlation
    and time fixed-effects, which seem to challenge the method of \cite{Qian2014}.\\
    Table \ref{DGP6} presents the case of no jumps (DGP6). The goal of examining DGP6 is
    to test whether our SAW estimation procedure is able to detect the no-jump case. The
    results from Table \ref{DGP6} show that the Post-SAW estimator is robust in this
    scenario. The small sample advantages of our method over the method of
    \cite{Qian2014} are again demonstrated in the simulation results.

\begin{table}[t!]
\setlength{\tabcolsep}{5pt}
\footnotesize\centering
\begin{tabular}{rr lll lll}
\toprule
\multicolumn{8}{c}{DGP5} \\
\midrule
&&\multicolumn{3}{c}{Post-SAW}&\multicolumn{3}{c}{\cite{Qian2014}}\\
\cmidrule(l{3pt}r{3pt}){3-5} \cmidrule(l{3pt}r{3pt}){6-8}
$T$&$n$&$\tilde{S}$&$||\hat{\beta}-\beta||^2/T$&HD$/T$&$\hat{S}$&$||\hat{\beta}-\beta||^2/T$&HD$/T$\\
\midrule
  33  & 30  &  3.000 (0.000) & 0.009 (0.006) & 0.000 (0.000) & 5.510 (1.696) & 0.037 (0.024) & 0.074 (0.062) \\ 
  33  & 60  &  3.000 (0.000) & 0.005 (0.003) & 0.000 (0.000) & 4.400 (1.237) & 0.015 (0.011) & 0.036 (0.037) \\ 
  33  & 120 &  3.000 (0.000) & 0.002 (0.001) & 0.000 (0.000) & 3.770 (0.855) & 0.006 (0.005) & 0.021 (0.026) \\ 
  33  & 300 &  3.000 (0.000) & 0.001 (0.001) & 0.000 (0.000) & 3.262 (0.523) & 0.002 (0.002) & 0.007 (0.014) \\ 
\midrule
  65  & 30  &  3.000 (0.000) & 0.005 (0.003) & 0.000 (0.000) & 6.176 (3.184) & 0.024 (0.020) & 0.072 (0.057) \\ 
  65  & 60  &  3.000 (0.000) & 0.002 (0.002) & 0.000 (0.000) & 4.110 (1.902) & 0.007 (0.008) & 0.031 (0.044) \\ 
  65  & 120 &  3.000 (0.000) & 0.001 (0.001) & 0.000 (0.000) & 3.356 (0.957) & 0.002 (0.003) & 0.012 (0.029) \\ 
  65  & 300 &  3.000 (0.000) & 0.000 (0.000) & 0.000 (0.000) & 3.026 (0.264) & 0.000 (0.001) & 0.001 (0.010) \\ 
\midrule
  129 & 30  &  3.000 (0.000) & 0.002 (0.002) & 0.000 (0.000) & 5.150 (3.469) & 0.011 (0.013) & 0.055 (0.072) \\ 
  129 & 60  &  3.000 (0.000) & 0.001 (0.001) & 0.000 (0.000) & 3.408 (1.154) & 0.002 (0.003) & 0.016 (0.042) \\ 
  129 & 120 &  3.000 (0.000) & 0.001 (0.000) & 0.000 (0.000) & 3.056 (0.370) & 0.001 (0.001) & 0.003 (0.018) \\ 
  129 & 300 &  3.000 (0.000) & 0.000 (0.000) & 0.000 (0.000) & 3.004 (0.089) & 0.000 (0.000) & 0.000 (0.005) \\ 
\bottomrule

\end{tabular}
\caption{Simulation results for DGP5 ($S=3$). The table shows the means (sd) computed
from the 500 Monte Carlo replications. Results are rounded to three decimal places.}%
\label{DGP5_S3}
\end{table}

The simulation results for the remaining DGPs (DGP2 and DGP3)
are essentially equivalent and, therefore, referred to Appendix C of the supplementary
paper \cite{Bada2018b}, which contains extended tables for DGP2-DGP5, showing all cases
$S=1,2,3$.

Overall, the Monte Carlo experiments show that, in many configurations of the data, our
method performs very well even in the case of endogeneity, presence of time
fixed-effects or when the idiosyncratic errors are affected by serial-autocorrelation
and/or heteroscedasticity, independently of the number and locations of the jumps, and
including small-$n$ scenarios.

\begin{table}[h!]
\small\centering
\begin{tabular}[c]{rr ll ll}
\toprule
\multicolumn{6}{c}{DGP6}\\
\midrule
&&\multicolumn{2}{c}{Post-SAW} & \multicolumn{2}{c}{\cite{Qian2014}}\\
\cmidrule(l{3pt}r{3pt}){3-4} \cmidrule(l{3pt}r{3pt}){5-6} $T$ & $n$ & $\tilde{S}$&$||\hat{\beta}-\beta||^2/T$ &$\hat{S}$&$||\hat{\beta}-\beta||^2/T$\\
\midrule
  33  & 30  & 0.000 (0.000) & 0.004 (0.006) & 7.490 (3.380) & 0.069 (0.031) \\ 
  33  & 60  & 0.000 (0.000) & 0.002 (0.002) & 3.672 (3.148) & 0.023 (0.017) \\ 
  33  & 120 & 0.000 (0.000) & 0.001 (0.001) & 1.054 (1.598) & 0.005 (0.007) \\ 
  33  & 300 & 0.000 (0.000) & 0.000 (0.000) & 0.090 (0.366) & 0.001 (0.001) \\ 
\midrule
  65  & 30  & 0.000 (0.000) & 0.002 (0.003) & 7.840 (5.329) & 0.045 (0.028) \\ 
  65  & 60  & 0.000 (0.000) & 0.001 (0.001) & 1.876 (2.958) & 0.008 (0.011) \\ 
  65  & 120 & 0.000 (0.000) & 0.000 (0.001) & 0.216 (0.692) & 0.001 (0.002) \\ 
  65  & 300 & 0.000 (0.000) & 0.000 (0.000) & 0.016 (0.141) & 0.000 (0.000) \\ 
\midrule
  129 & 30  & 0.000 (0.000) & 0.001 (0.001) & 2.758 (4.284) & 0.011 (0.016) \\ 
  129 & 60  & 0.000 (0.000) & 0.000 (0.001) & 0.200 (0.705) & 0.001 (0.002) \\ 
  129 & 120 & 0.000 (0.000) & 0.000 (0.000) & 0.028 (0.244) & 0.000 (0.001) \\ 
  129 & 300 & 0.000 (0.000) & 0.000 (0.000) & 0.004 (0.063) & 0.000 (0.000) \\ 
\bottomrule

\end{tabular}
\caption{Simulation results for DGP6 ($S=0$). The table shows the means (sd) computed
from the 500 Monte Carlo replications. Results are rounded to three decimal places.}%
\label{DGP6}%
\end{table}


\section{Application: Algorithmic Trading and Market Quality} 

\label{application} As an empirical vehicle for our estimation methodology, we examine the, potentially, time varying effects of high frequency algorithmic trading (AT) on measures of market quality.

A number of studies (see, for example \citet{Hendershott2011} and
\citet{Boehmer2012}) have found a positive relationship between AT and measures
of liquidity in financial markets. However, incidences such as the \quotes{Flash
  Crash}, albeit anecdotal, suggest a possible time varying relationship between
the two, whereby AT increases liquidity under \quotes{normal} market conditions
but may lead to undesirable outcomes during periods of stress. For example, \cite{Boneva2016} document that dark trading increases the
volatility of certain measures of market quality using data from the UK equity
markets, and \cite{Linton_2018} survey a number of empirical studies that
document similar outcomes during selected stressed periods in various markets.
Regulators in the US and Europe have also given the effects of AT increased
scrutiny and proposed new market rules designed to track and monitor the
positions taken by AT firms (see \citet{conghft} for an extensive discussion).

Of particular importance to the concept of liquidity is the timing of its provision. The merits of added liquidity during stable market periods at the expense of its draw back during periods of higher uncertainty can potentially leave investors worse off, and it is exactly this situation that high-frequency trading firms can exploit and apparently did exploit during the early years of the Great Recession.  The model of \citet{Acharya2005} is particularly relevant as it exemplifies the many avenues through which time varying liquidity can affect expected returns. They show that asset returns are increasing in the covariance between portfolio illiquidity and market illiquidity and decreasing in the covariance between asset illiquidity and the market return. A consequence of this is that if AT intensifies these liquidity dynamics for a particular asset then the effect will be to increase the risk premium associated with that security, which represent higher costs of capital for firms.  Thus increased AT can potentially decrease firm investment through its effect on liquidity dynamics and we are able to identify with our new modeling approach periods during which the AT and market quality relationship changed abruptly.

\subsection{Data}
Our dataset is made up of a balanced panel of stocks whose primary exchange is the New York Stock Exchange (NYSE) and covers the time period $2003-2008$. To build our measures of market quality, we use the NYSE Trade and Quotation (TAQ) database provided by Wharton Research Data Services (WRDS) to collect intra-day data on stock trades. We construct daily liquidity measures and then average them over the course of the calendar month. The full sample consists of 378 firms over 71 months. In specifications where the parameters are allowed to vary we choose the latest 64 months to meet the requirements of our estimation approach.  We combine our high-frequency measures of liquidity with information on price and shares outstanding from the Center for Research in Security Prices (CRSP), also available via WRDS.

\subsubsection{The Algorithmic Trading Proxy}
Our AT proxy is motivated by \citet{Hendershott2011} and \citet{Boehmer2012}, who note that AT is generally associated with an increase in order activity at smaller dollar volumes. Thus, the proxy we consider is the negative of dollar
volume (in hundreds of dollars, \$Vol$_{it}$) over time period $t$ divided by total order activity over time period $t$. We define order activity as the sum of trades (Tr$_{it}$) and updates to the best prevailing bid and offer ($q_{it}$) on the security's primary exchange:

\[
\text{AT}_{it} = - \frac{\text{\$Vol}_{it}}{\text{Tr}_{it} + q_{it}}.
\]

An increase in AT$_{it}$ represents a decrease in the average dollar volume per instance of order activity and represents an increase in the AT in the particular security.

\subsubsection{Market Quality Measures}
We consider several common measures of market quality to assess the impact of AT on markets for individual securities.

\subsubsection*{Proportional Quoted Spread}
The proportional quoted spread (PQS$_{it}$) measures the quoted cost, as a percentage of the price (Bid-Offer midpoint), of executing a trade in security $i$ and is defined as,
\[
\text{PQS}_{it}=100\left(  \frac{Ofr_{it}-Bid_{it}}{0.5(Ofr_{it}+Bid_{it}%
)}\right)  .
\]
We multiply by $100$ in order to place this metric in terms of percentage points. We aggregate this metric to a monthly quantity by computing a share volume-weighted average over the course of each month.

\subsubsection*{Proportional Effective Spread}
The proportional effective spread (PES$_{it}$) is quite similar to (PQS$_{it}$) but accommodates potentially hidden liquidity or stale quotes by evaluating the actual execution costs of a trade. It is defined as,
\[
\text{PES}_{it}=100\left(  \frac{|P_{it}-M_{it}|}{M_{it}}\right)  ,
\]
where $P_{it}$ is the price paid for security $i$ at time $t$ and $M_{it}$ is the midpoint of the prevailing bid and ask quotes for security $i$ at time $t$.  We again aggregate this measure up to a monthly quantity in the same way as we do for quoted spreads.

\subsubsection*{Measures of Volatility}
We also consider two different measures of price volatility in security $i$ over time period $t$. The first is the daily high-low price range given by,
\[
\text{H-L}_{it} = 100 \left(  \frac{\max_{\tau \in t}(P_{it}) - \min_{\tau \in
t}(P_{it})}{P_{it}} \right)  ,
\]
which represents the extreme price disparity over the course of a trading day. We also consider the realized variance of returns over each day computed using log percentage (i.e. $\ln(p_{i,t}/p_{i,t-1})\times100$) returns over
$5$-minute intervals:
\[
\text{RV}_{it} = \left(  \sum_{\tau \in t}r^{2}_{i\tau} \right)  .
\]
Realized variance is a non-parametric estimator of the integrated variance over the course of a trading day (see, for example, \citet{andersen2003modeling}). We aggregate both measures up to a monthly level by averaging over the entire
month. We additionally represent each in terms of percentages, i.e. $H-L_{it}$ is the price range as a percentage of the daily closing price and $RV_{it}$ is an estimate of the integrated variance of log returns in percentages.

\subsubsection{Additional Control Variables}
We also include control variables, lagged by one month so they are
representative of the beginning state of the market. The control variables are:
(1) Share Turnover (ST$_{it}$), which is the number of shares traded over the
course of a day in a particular stock relative to the total amount of shares
outstanding; (2) Inverse price, which represents transaction costs due to the
fact that the minimum tick size is $1$ cent; (3) Log of the market value of equity to accommodate effects associated with smaller
securities; (4) Daily price range to accommodate any effects from large price swings in the previous month. To avoid adding lagged dependent variables in the
model, for regressions where the daily price range is the dependent variable we
replace it in the vector of controls with the previous month's realized variance. We additionally include security and time period fixed
effects to proxy for any time period or security related effects not captured by
our included variables.

The control variables selected in our specification have been investigated
substantially in the literature, both theoretically and empirically \citep[for a broad
survey the reader is referred to][]{vayanos2013market}. We expect, at least
directionally, for their impact to be consistent with this literature.
Therefore, in this case we focus solely on the potential for breaks in the
parameter associated with our AT proxy, for which the evidence remains
inconclusive, and assume parameters associated with all other control variables
are constant over the sample period.

\subsubsection*{Potential Endogeneity Issue}
We assume a linear relationship between our measures of market quality, our proxy for AT and our control variables,
\begin{equation}
\label{empmodel}\text{MQ}_{it}=\mu+\alpha_{i}+\gamma_{t}+\text{AT}_{it}\beta_{t}%
+X_{it}^{\prime}\delta+\epsilon_{it}.
\end{equation}
Absent a theoretical model of AT, an issue on which the literature is still somewhat agnostic, it is uncertain whether AT strategies attempt to time shocks to market quality. This creates a potential problem of endogeneity with our AT proxy. That is, when estimating the regression equation \eqref{empmodel} our estimates may be biased ($E(\text{AT}_{it}%
e_{it})\neq0$) and inconsistent.

To overcome this potential issue we use the approach of \citet{Hasbrouck2013} (albeit with different variables) and choose as an instrument the average value of AT over all other firms not in the same industry as
firm $i$. To this end, we define industry groups using $4$-digit SIC codes and define these new variables as AT$_{-IND,it}$. The use of this instrument requires some commonality in the level of AT across all stocks that is sufficient to pick up exogenous variation. It further rules out trading strategies across firms in different industry groups. Lacking much
knowledge of the algorithms used by AT firms we view this assumption as reasonable. As noted by \citet{Hasbrouck2013}, it is unlikely that AT firms implement cross-stock trading strategies for a particular firm with the entire universe of firms (or in our case the other $377$). To the extent that AT firms do implement these cross-stock strategies across industries, their
effect on the average is likely to be marginal.

To estimate the model we use a two-stage approach and first fit the regression model,
\begin{equation}
\label{simultanequ}\text{AT}_{it} = a_{i} + g_{t} + b\text{AT}_{-IND, it} +
dW_{it} + \epsilon_{it}%
\end{equation}
to obtain an instrument, $Z_{it}$, for AT$_{it}$ given by the fitted values from \eqref{simultanequ}, i.e., $Z_{it} := \hat{\text{AT}}_{it} = \hat{a}_{i} + \hat{g}_{t} + \hat{b}\text{AT}_{-IND, it} +\hat{d} W_{it} $, where $\hat{a}_{i} $, $\hat{g}_{t}$, $\hat b$, and $\hat d$ are the conventional estimates of ${a}_{i}$, ${g}_{t}$, $b$, and $d$. We then carry out the second stage regression using equation \eqref{empmodel} using the estimator discussed in Section 3.3. For comparison purposes, we additionally apply the conventional panel data model assuming a constant slope parameter, i.e.,
$\beta_{1} = \beta_{2} = \cdots= \beta_{T}$.

\begin{table}[tb!]
\begin{center}
\begin{tabular}{lc  cccccc}
\toprule	
Dependent&&\multicolumn{6}{c}{Regressors}\\
Variable&  & $\hat{\textrm{AT}}_{it}$ & $\ln(\textrm{ME})_{i,t-1}$& $\textrm{T}/\textrm{O}_{i,t-1}$ & $1/\textrm{P}_{i,t-1}$ & $\textrm{H-L}_{i,t-1}$& $\textrm{RV}_{i,t-1}$\\ \hline
PQS$_{it}$ & Coef. & -0.013 & -0.003 & 0.027 & 0.619 & 0.002  & \\
& $t$-value  & -3.61 & -2.65 & 0.73 & 15.38 & 5.67 &  \\
\midrule
PES$_{it}$ & Coef. & -0.006 & -0.001 & -0.077 & 0.517 & 0.004 & \\
 & $t$-value  & -3.62 & -1.83 & -3.42 & 18.51 & 16.96 & \\
 \midrule
RV$_{it}$ & Coef. & -0.691 & 0.046 & -1.575 & 7.25 & 0.415 & \\
 & $t$-value  & -12.73 & 2.39 & -2.45 & 10.19 & 44.46  & \\
 \midrule
 H-L$_{it}$ & Coef. & -1.404 & -0.038 & -6.15 & 7.88 & & 1.151 \\
 & $t$-value  & -11.6 & -1.04 & -4.71 & 6.04 & & 35.78 \\
 \bottomrule
\end{tabular}
\caption{IV panel data model with constant parameters. For the purpose of readability, we divide the AT variable by 100 to reduce trailing zeros after the decimal.  The table shows the results of the 2SLS panel regression
of our measures of market quality on our AT proxy and further regressors, namely, the previous month's log of market Cap ($\ln(\text{ME})$), share turnover ($\text{T}/\text{O}$), inverse price ($1/\text{P}$) and high-low price range $(\text{H-L})$. When the dependent variable is the current month's high-low price range, last
month's value of realized variance ($\text{RV}$) is used instead to avoid a
dynamic panel model. Standard errors are corrected for heteroscedasticity.}
\label{convmod}%
\end{center}
\end{table}

\subsection{Results}
Table \ref{convmod} presents the results from a baseline model that assumes the slope parameters are constant over time. These results are largely consistent with previous studies that find a positive (in terms of welfare) average relationship between AT and measures of market quality over the time period considered. The coefficient
estimates on the AT proxy are negative and significant for all four measures of market quality that we consider. That is, increases in AT generally reduce both of the spread measures and both of the variance measures we consider. As
for the direction of the effect, differences in our proxies and choice of instruments do not seem to reach conclusions that are at odds with the prior literature.

\begin{figure}[!h!]
  \centering
  \subfigure[Proportional quoted spread (PQS).]
  {\includegraphics[width=.49\textwidth]{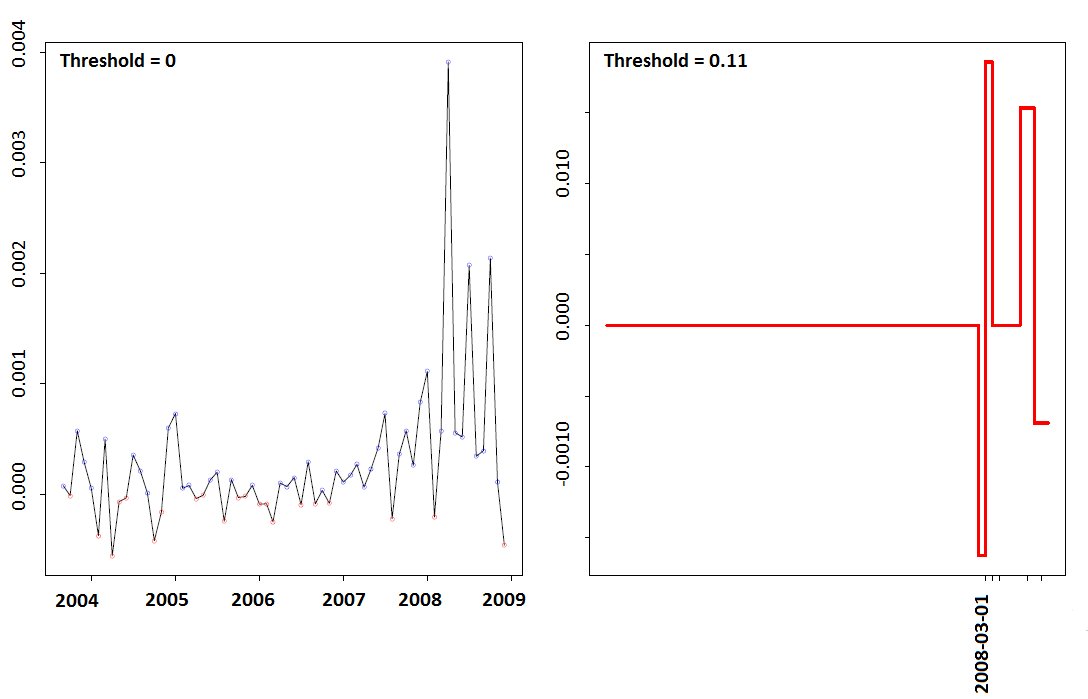}}
  \subfigure[Proportional effective spread (PES)]
  {\includegraphics[width=.49\textwidth]{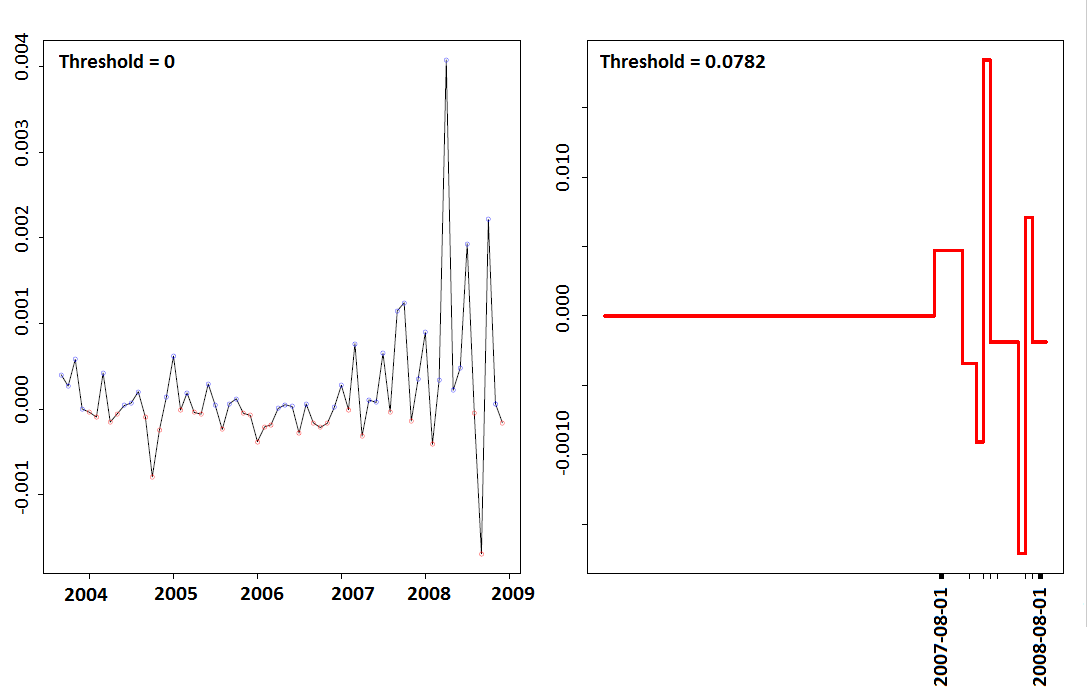}}
  \subfigure[High-low price range (H-L)]
  {\includegraphics[width=.49\textwidth]{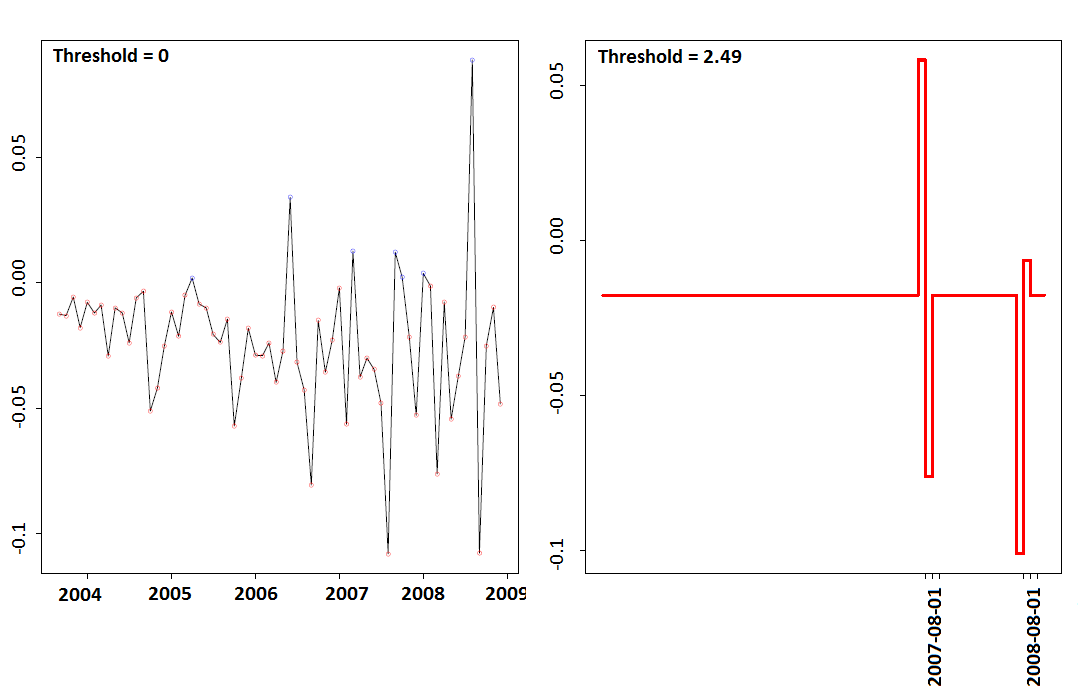}}
  \subfigure[Realized variance (RV)]
  {\includegraphics[width=.49\textwidth]{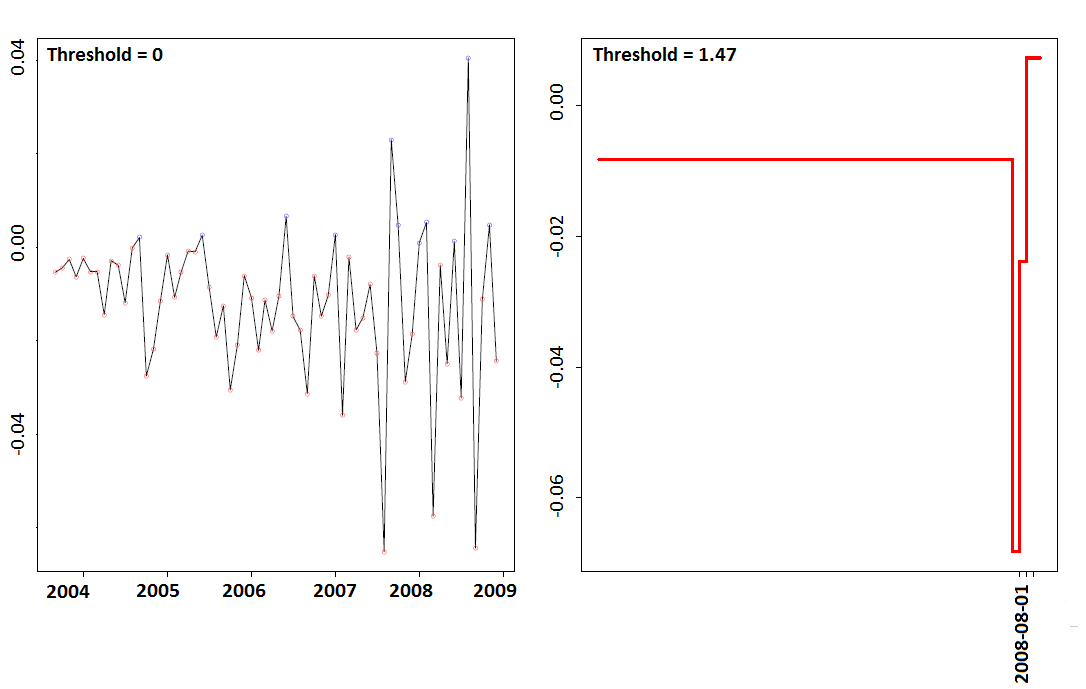}}  
  \caption{Time
    varying effects of algorithmic trading on different outcomes (PQS,
    PES, H-L, and RV). The left plots show the estimated post-SAW
    coefficients and the right plots show the results from naive period by
    period cross-sectional regressions.}\label{fig:Appl}
\end{figure}

Table \ref{ResultTab}  presents the results when we
allow the parameter to jump discretely over time. For each of the detected time
  intervals (``stability intervals''), defined by the
  estimated jump-locations, we report the post-SAW coefficient
  estimates. To test the statistical significance of the
  difference of two consecutive coefficient estimates, we use a classic Chow test based on our asymptotic
  normality result in Theorem \ref{theo4}. Figure \ref{fig:Appl} (a)-(d) show
the estimated post-SAW coefficients (right plots) and the results from period by
period cross-sectional regressions (left plots).

As is quite evident from the plots of the post-SAW coefficients, for a large portion of our sample time period we find a stable relationship between AT and our measures of market quality. This was a relatively placid period for equity markets and the lack of time varying effects accords with our prior belief that structural breaks in the marginal effect are likely to occur during periods of turmoil.

During the financial crisis period, loosely defined here as the 2007-2008 time period, we find a significant evidence of breaks in the coefficient on our AT proxy. For the two spread measures we find evidence of large positive jumps in the coefficients in April and September/October of 2008. During these two months an increase in AT lead to an increase in spreads, counter to the average findings of Table \ref{convmod}. During this period, transacting in stocks with high AT was, other things being equal,  costlier than in low AT securities. We also find similar breaks in the relationship between AT and our variance measures during the same time period. Of note is that for realized variance we find jumps to be beneficial for investors. That is, we find that increases in AT cause a larger reduction in realized variances.

\begin{table}[h!]
  \begin{center}
    \begin{tabular}{llccc}
      \toprule
      Dependent&&                     & \multicolumn{2}{c}{Chow-Test}         \\
      Variable&Stability Intervals  & Coef.					     &$z$-score &$p$-value  \\
      \midrule
      PQS$_{it}$&from 2003-09-01 to 2008-02-01&\phantom{$-$}6.49e-05&  	             & 		       \\
      &from 2008-03-01 to 2008-03-01&\phantom{$-$}6.51e-04&\phantom{$-$}1.65&\phantom{$<$}0.10\\
      &from 2008-04-01 to 2008-04-01&\phantom{$-$}4.13e-03&\phantom{$-$}6.42&$<$0.01\\
      &from 2008-05-01 to 2008-08-01&\phantom{$-$}7.66e-04&$-$7.42          &$<$0.01\\
      &from 2008-09-01 to 2008-10-01&\phantom{$-$}1.03e-03&\phantom{$-$}0.93&\phantom{$<$}0.35   \\
      &from 2008-11-01 to 2008-12-01&$-$1.46e-04          &$-$4.62          &$<$0.01\\ 
      \midrule
      PES$_{it}$&from 2003-09-01 to 2007-08-01  &\phantom{$-$}9.06e-06   &\phantom{$-$}		& 		\\
      &from 2007-09-01 to 2007-12-01  &\phantom{$-$}6.73e-04   &\phantom{$-$}3.75   &\phantom{$<$}0.01   \\
      &from 2008-01-01 to 2008-02-01  &\phantom{$-$}1.35e-04   &$-$2.00  &$<$0.05   \\
      &from 2008-03-01 to 2008-03-01  &\phantom{$-$}5.31e-04   &\phantom{$-$}1.16   &\phantom{$<$}0.25   \\
      &from 2008-04-01 to 2008-04-01  &\phantom{$-$}4.19e-03   &\phantom{$-$}10.70  &$<$0.01\\
      &from 2008-05-01 to 2008-08-01  &\phantom{$-$}4.15e-04   &$-$15.60 &$<$0.01\\
      &from 2008-09-01 to 2008-09-01  &$-$1.74e-03             &\phantom{0}$-$7.54  &$<$0.01\\
      &from 2008-10-01 to 2008-10-01  &\phantom{$-$}1.79e-03   &\phantom{$-$}11.70  &$<$0.01\\
      &from 2008-11-01 to 2008-12-01  &\phantom{$-$}3.55e-06   &\phantom{$-$}$-$9.61  &$<$0.01\\ 
      \midrule
      H-L$_{it}$&from 2003-09-01 to 2007-06-01&$-$0.017&     &     \\
      &from 2007-07-01 to 2007-07-01&$-$0.049&$-$1.08&\phantom{$<$}0.29\\
      &from 2007-08-01 to 2007-08-01&$-$0.154&$-$2.68&\phantom{$<$}0.01\\
      &from 2007-09-01 to 2008-08-01&$-$0.012&\phantom{$-$}5.23   &$<$0.01\\
      &from 2008-09-01 to 2008-09-01&$-$0.107&$-$4.40  &$<$0.01\\
      &from 2008-10-01 to 2008-10-01&$-$0.001&\phantom{$-$}4.59   &$<$0.01\\
      &from 2008-11-01 to 2008-12-01&$-$0.021&$-$1.60  &\phantom{$<$}0.11\\
      \midrule
      RV$_{it}$&from 2003-09-01 to 2008-08-01  &$-$0.008   &$-$14.20 &$<$0.01\\
      &from 2008-09-01 to 2008-09-01  &$-$0.063   &$-$5.09  &$<$0.01\\
      &from 2008-10-01 to 2008-10-01  &\phantom{$<$}0.001   &\phantom{$-$}5.29  &$<$0.01\\
      &from 2008-11-01 to 2008-12-01  &$-$0.008   &$-$1.26&\phantom{$<$}0.21\\
      \bottomrule
    \end{tabular}
    \caption{The column ``Coef.'' shows the post-SAW estimates for each stability intervals defined by the
      estimated jump-locations.  The (classic) Chow test tests for the significance
      of the parameter change between two consecutive time periods.}%
    \label{ResultTab}
  \end{center}
\end{table}

\section{Conclusion}

\label{conclus} This paper generalizes existing panel model specifications in
which the slope parameters are either constant over time or display time
heterogeneity. We allow for multiple structural changes that can occur at
unknown date points and may affect each slope parameter separately. Consistency
under weak forms of dependency and heteroscedasticity in the idiosyncratic
errors is established and convergence rates are derived. Our empirical
illustration finds evidence that the relationship between high frequency
algorithmic trading and market quality was disrupted during the time between
$2007$, beginning of the subprime crisis in the US market, and $2008$, the
bankruptcy of financial services firm Lehman Brothers.

\paragraph{Acknowlegements}
Previous versions of this paper were given at the 15th International Symposium
on Econometrics, Operations Research and Statistics, Süleyman Demirel
University, Isparta, Turkey; The 3rd International Workshop Measuring Banking
Performance, University of Loughborough, UK, June, 2014; the 6th International
Finance and Banking Society (IFABS) Conference on Alternative Futures for Global
Banking: Competition, Regulation and Reform, Lisbon, June 2014; University of
Sydney, Department of Business Analytics Seminar, August, 2014: School of
Economics, University of Queensland, Economics Seminar Program, October, 2014;
Monash University Department of Econometrics and Business Statistics, November,
2014; School of Economics Seminar in Econometrics, Georgia Institute of
Technology, April, 2015; Conference on Econometric Methods for Banking and
Finance, Banco de Portugal, September, 2014; New York Camp Econometrics X,
April, 2015; Econometric Society 2015 World Congress, August, 2015; Michigan
State University, December, 2015.  The authors would like to thank participants
of those conferences and seminars for their helpful comments and criticisms. We are especially grateful to the four anonymous referees and the editors
for their valuable and constructive comments which helped to improve our
manuscript.

\paragraph{Supplementary Material}
\begin{itemize}
\item The estimation methods introduced in this paper are freely available in our
  \textsf{R}-package \texttt{sawr} which can be downloaded and installed from:\\
  \url{https://github.com/timmens/sawr/}.
\item The simulation results can be fully reproduced using the \textsf{R}-codes at:\\
  \url{https://github.com/timmens/simulation-saw-paper}.
\item All proofs and additional
  theoretical discussions and simulation results can be found in our supplementary paper \cite{Bada2018b}.
\end{itemize}

\bibliography{library}
\bibliographystyle{apalike}

\end{document}